\documentclass[acmsmall]{acmart}

\AtBeginDocument{%
  \providecommand\BibTeX{{%
    \normalfont B\kern-0.5em{\scshape i\kern-0.25em b}\kern-0.8em\TeX}}}

\setcopyright{none}
\acmYear{}
\acmDOI{}

   \acmConference[Facebook AI]{Facebook AI}{September}{2020}
  
\acmPrice{}
\acmISBN{}

\begin{document}

\title{Preserving Integrity in Online Social Networks}


\author{Alon Halevy}
\affiliation{
\institution{Facebook AI}
}
\email{ayh@fb.com}
\author{Cristian Canton-Ferrer}
\affiliation{
\institution{Facebook AI}
}
\author{Hao Ma}
\affiliation{
\institution{Facebook AI}
}
\author{Umut Ozertem}
\affiliation{
\institution{Facebook AI}
}
\author{Patrick Pantel}
\affiliation{
\institution{Facebook AI}
}
\author{Marzieh Saeidi}
\affiliation{
\institution{Facebook AI}
}
\author{Fabrizio Silvestri}
\affiliation{
\institution{Facebook AI}
}
\author{Ves Stoyanov}
\affiliation{
\institution{Facebook AI}
}




\renewcommand{\shortauthors}{Halevy, Canton-Ferrer, et al.}
\begin{abstract}
   Online social networks provide a platform for sharing information and free expression. However, these networks are also used for malicious purposes, such as distributing misinformation and hate speech, selling illegal drugs, and coordinating sex trafficking or child exploitation. This paper surveys the state of the art in keeping online platforms and their users safe from such harm, also known as the problem of preserving integrity.  
    
     This survey comes from the perspective of having to combat a broad spectrum of integrity violations at Facebook. We highlight the techniques that have been proven useful in practice and that deserve additional attention from the academic community.
     Instead of discussing the many individual violation types, we identify key aspects of the social-media eco-system, each of which is common to a wide variety violation types. Furthermore, each of these components represents an area for research and development, and the innovations that are found can be applied widely.
\end{abstract}



\maketitle

\section{Introduction}
\par The goal of online social networks is to help create connections between people (online and offline), to connect people to communities of interest, and to provide a forum for advancing culture. Social networks advance these causes by providing a platform for free expression by anyone, be they well-known figures or your next door neighbor. Unfortunately, open platforms for free expression can be used for malicious purposes. People and organizations can distribute misinformation and hate speech, and can use the platform to commit crimes such as selling illegal drugs, coordinating sex trafficking or child exploitation. All of these violations existed much before the advent of social networks, but social networks exacerbate the scale and sophistication with which these activities can be carried out. 

Naturally, fighting these violations, which we collectively refer to as the problem of {\em preserving integrity in online networks} (or simply, integrity), has become a huge priority for the companies running them and for society at large. 
The challenges in preserving integrity fall into two general categories: policy and technical. Setting policies for what content and behavior are allowed on social networks is an area fraught with debate because it involves striking a balance between free expression and removing offending content. In addition, the policies need to be sensitive to a variety of cultures and political climates all over the world. While we touch on the policy backdrop, this survey focuses on the technical challenges that arise in enforcing the policies. Fundamentally, the technical challenges arise because deciding whether a post is violating can be extremely subtle and depends on deep understanding of the cultural context. To make things worse, content (violating and non-violating) is created at unprecedented scale,  in over  100 languages, and in very differing norms of social expression.  In addition, preserving integrity is a problem with an adversarial nature--as the actors learn the techniques used to remove violating content, they fine tune their techniques to bypass the safeguards. 

The academic community has been actively researching integrity problems for the past few years and several surveys have been written about specific aspects of the general problem of integrity (e.g.,~\cite{DBLP:journals/corr/abs-1902-07539,DBLP:journals/pvldb/LakshmananST19,DBLP:journals/tist/SharmaQJRZL19}).  This survey comes from the perspective of having to combat a broad spectrum of integrity violations at Facebook. The problems that Facebook has had to tackle have also been experienced on other social networks to varying degrees.\footnote{While this survey is based on our experience at Facebook, it is not meant to be a description of how Facebook tackles integrity or Facebook's stance on integrity issues. When we use examples of policies or systems used at Facebook, we call them out explicitly.} The breadth of the services that Facebook offers, the variety of the content it supports and the sheer size of its user base have likely attracted a widest set of integrity violations, and in many cases, the fiercest. This survey does not attempt to cover each type of integrity violation. Instead, we identify a few sets of techniques that together form a general framework for addressing a broad spectrum of integrity violations, and highlight the most useful techniques in each category. 

Admittedly, despite its huge importance to society, integrity is a tricky area for collaboration between industry and academia. One reason is that sharing datasets can be problematic. In addition to the usual concerns involved in sharing confidential user data, some data is simply illegal to share (e.g., a dataset of child exploitation imagery). The second reason is that some of the methods that are used to preserve integrity need to be kept confidential, otherwise they can be weaponized by bad actors. To address some of these issues, companies have invested significant effort to create datasets that can be used for research purposes, such as the Deepfake Detection Challenge~\cite{Dolhansky2019TheDD} and the Hateful Memes Dataset~\cite{kiela-memes}. This survey highlights the techniques that have been found useful in practice  and therefore merit additional academic research that can contribute indirectly to combating integrity.

The survey is structured as follows. Section~\ref{section:definition} defines the problem of integrity and sets the context for how content flows through the integrity enforcement mechanisms of social networks. Section~\ref{section:framework} describes a holistic view of the work on integrity that is inspired by the plethora of current work and the challenges we face in practice. Section~\ref{section:content} describes techniques for analysing the contents of a social-media post, and Section~\ref{section:behavior} addresses how integrity algorithms can leverage the interactions that users and other actors have with the post. Section~\ref{section:others} describes emerging challenges in the area of preserving integrity.

\section{Problem definition}
\label{section:definition}
The problem of preserving integrity in social networks is very broad. We begin this section by defining the integrity problem as one of ensuring that content uploaded to social networks and the behavior of its users adheres to a set of policies put forth by the companies running the services. We then explain the typical workflow of integrity management employed by social networks thereby highlighting where technology plays a role. Finally, we turn to the issue of measuring the success of techniques for preserving integrity.  

\subsection{Social network basics}
\label{subsection:basics}
Social networks vary widely in the features and interactions they offer. However, the following concepts are common across them. The basic type of content on a social network is a post. A post can be made by an individual or by an entity such as a company or an interest group. Posts may be public, or visible only to friends of the individual or to a group. In some cases, groups are closed in that they require approval by an owner to join. Content posted as described above is referred to as {\em organic content}. In contrast, ads that are put on users' streams are {\em paid content}. The vast majority of ads are commercial (i.e., posted by a company trying to sell products). Ads about social issues, elections or politics are a category of ads on important topics such as political figures, ``get out the vote'' campaigns, or ads that take a stand on sensitive issues, like health and the environment. These topics are heavily debated, can influence many people and may influence the outcome of an election, or result in or relate to existing or proposed legislation. These ads are scrutinized further. For example, to run these ads, Facebook requires advertisers go through the authorization process to confirm their identity and location, place ``Paid for by'' disclaimers on ads and have their ads enter the Ad Library for seven years.

Social networks typically offer messaging platforms where users can interact with members of their social networks. Messaging has also been a vehicle for violating content policies, such as grooming children for exploitation at a later time.  As messaging services move towards end-to-end encryption, social networks need to find the right balance between privacy offered by the encryption and safety which can be further violated when messages are private. 
We discuss some of these issues in Section~\ref{section:privacy}.

\subsection{Policies}
\label{sec:violations}

\begin{table}[ht]
\begin{center}
\begin{tabular}{| p{0.15\textwidth} | p{0.4\textwidth} | p{0.35\textwidth} |}
\hline
{\bf Category} &
{\bf You cannot...} &
{\bf But you can...} \\ \hline 
Violence and criminal behavior &
\begin{footnotesize} 
\begin{itemize}
\item have a presence on FB if you have a violent mission
\item support individuals or organizations with violent missions,
\item incite violence, or admit to violent activities.
\end{itemize}
{\em Violent activities include: terrorism, organized hate, mass murder, human trafficking, organized violence or criminal activity}.
\end{footnotesize} &
\begin{footnotesize}
\begin{itemize}
\item provide instructions on how to create or use explosives if there is clear context that it is for non-violent purposes (e.g.,  part of commercial video games, clear scientific/educational purpose, fireworks, or specifically for fishing).
\end{itemize}
\end{footnotesize}
\\ \hline

Illegal trade & 
\begin{footnotesize}
\begin{itemize}
\item sell firearms or illegal drugs or marijuana
\item post content that is meant to deliberately deceive people or deprive them of money, legal rights or property.
\end{itemize}
\end{footnotesize} &

\begin{footnotesize}
\begin{itemize}
 \item debate crimes and the laws governing sales of certain merchandise.
 \item gun retailers can advertise what is available for sale in their stores.
\end{itemize}
\end{footnotesize}
\\ \hline

Safety &
\begin{footnotesize} 
\begin{itemize}
\item encourage suicide or bodily damage
\item negatively target survivors of suicide or self injury 
\item post child nudity or exploitation of children 
\item depict, threaten or promote sexual exploitation of adults
\item post content that depicts or promotes sexual acts with non-consenting adults 
\item bully people (threat, message, release personal information, degrade, shame, e.g., by discussing their sexual activity) 
\item promote or depict exploitation of humans (e.g., human trafficking, human smuggling). 
\item post private information about others without their consent. 
\end{itemize}
\end{footnotesize} 
&
\begin{footnotesize}
\begin{itemize}
\item if there is a chance to help individuals who post content about injuring themselves then the content may stay up while there is a chance they can be helped. 

\item discuss the issues of sexual exploitation and victims can share their experiences. 

\item discuss issues of concerning experiences such as suicide, bodily damage, and sexual exploitation, and victims of such crimes can share their experiences. 

\item public figures are treated differently to allow discussion of them.
\end{itemize}
\end{footnotesize}
\\ \hline 
\end{tabular}
\end{center}
\caption{Examples summarized from the Facebook Community Standards\protect~\cite{facebook-standards}. }
\label{table:violations}
\end{table}

The problem of preserving integrity on social networks is defined by the community policies published by these networks that describe what is allowed on their platforms (e.g., the Twitter Rules~\cite{twitter-standards}, YouTube Community Guidelines~\cite{youtube-standards}, The Facebook Community Standards~\cite{facebook-standards}). While the final formulation of the policies is determined by the companies themselves, they are based on significant input from the community (e.g., the European Commission~\cite{europe-standards}) and local laws.  The challenge in setting these policies is to balance free expression with the desire to keep the platform safe. For example, posting a bloody body is likely not allowed. However, if the context is a birth scene then it could be allowed, as long as it's not showing private body parts. As we illustrate later, simple policies prohibiting certain content typically have ramifications of preventing valuable content by people whose voice would otherwise not be heard. It is also important to note that these policies are not static. The policies and enforcement guidelines are updated as online discourse changes and new nefarious uses of social networks arise.
For example, when a new type of misinformation surfaced that may result in  physical harm  (e.g., bogus treatments against COVID-19~\cite{vaccine-misinfo-blog}), reviewer guidance was updated to be clear, a policy to remove content leading to imminent physical harm applied to this content.

Tables~\ref{table:violations} and~\ref{table:violations1} provide examples that are summarized from the Facebook Community Standards~\cite{facebook-standards} as to what is not allowed to post (second column), but also closely related content that may be allowed (third column). There are a few points to note about these policies.

\begin{table}[hbt]

\begin{center}

\begin{tabular}{| p{0.15\textwidth} | p{0.4\textwidth} | p{0.35\textwidth} |}
\hline
{\bf Category} &
{\bf You cannot...} &
{\bf But you can...} \\ \hline 

Objectionable content &
\begin{footnotesize} 
\begin{itemize}
\item post hate speech,  defined as content that attacks people based on their race, ethnicity, national origin, religious affiliation, sexual orientation, caste, sex, gender, gender identity, and serious disease or disability

\item post content that glorifies the suffering or humiliation of others 
\item post content involving sexual solicitation
\item post photos depicting nudity.
\end{itemize} 
\end{footnotesize} &

\begin{footnotesize}
\begin{itemize}
\item some of the same words that are used in hate speech may be used in contexts where the intent is not hate speech and those are allowed
\item humor and social commentary about these subjects is allowed 
\item content whose intent is of raising the issues of hate speech are allowed 
\item posts that are meant to bring attention to sexual solicitation are allowed
\item some content may be allowed but labeled as sensitive for younger users 
\item some sexual imagery is allowed if it's for educational or satirical purposes.
\end{itemize}
\end{footnotesize}
\\ \hline 
Authenticity &
\begin{footnotesize} 
\begin{itemize}
\item {\em should} represent yourself using your real-life name 
\item post spam or try to artificially increase viewership for commercial gain
\item attempt to gather private information
\item create fake accounts
\item create and distribute false news
\item create manipulated media with the intent to deceive. 
\end{itemize}
\end{footnotesize} &

\begin{footnotesize}
There is a fine line between false news and opinion, so false news is demoted, not necessarily removed. \end{footnotesize}
\\ \hline 
Intellectual property &
\begin{footnotesize}  misuse intellectual property rights.  
\end{footnotesize}& \\
\hline

\end{tabular}
\end{center}
\caption{A continuation of Table~\ref{table:violations}.}
\label{table:violations1}
\end{table}

First, while these policies provide guidelines, it is very hard to set hard and fast rules. For example, while it is easy to  agree that posting nude photos of children is objectionable, there are still rare counterexamples where it has cultural value, such as the photo of the 9 year old girl in Saigon in 1972 that won the Pulitzer Prize~\cite{saigon-photo}. In other cases, the {\em intent} of the post is crucial to determining whether it is violating or not. For example, it often happens that certain derogatory words are {\em adopted} for good intent by their target community, and therefore their use is benign. The intent of a post is often subjective and very hard to determine even for human inspectors, let alone algorithms.

The second point to note is that some violations are described in terms of content (e.g., nudity or hate speech), whereas others concern the users' behavior on the social network. One aspect of behavior is the identity one presents on the social network such as impersonations. A subtle example of impersonation is when groomers for child abuse join groups that attract kids. Another aspect of behavior involves coordinated activities of multiple parties. For example, account farms are used to distribute voter suppression campaigns. As we see later in the paper, the distinction between content and behavior based violations is also pertinent to the techniques developed for catching them. Whereas many techniques focus on the content itself, it turns out that targeting the network effects of bad behavior can often provide significant leverage in tackling integrity problems. 

Finally, there is a large disparity in the frequency of violations. Hate speech and nudity occur relatively frequently, whereas child abuse and trafficking is relatively rare. As a result, the amount of training data that is available for each kind of violation varies widely and measuring effectiveness of the enforcement solutions is tricky.

\subsection{Enforcing integrity}

\begin{figure}[ht]
\begin{center}

\includegraphics[width=5in]{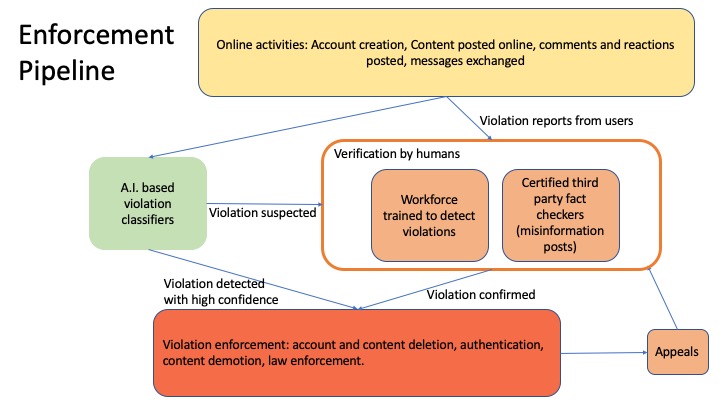}
\caption{The flow of content in a social network. As content is uploaded to the network, it is inspected by integrity classifiers. Content that the classifier deems as violating with very high confidence may be removed immediately. Content that is deemed potentially violating by the classifiers is sent to be reviewed by trained content reviewers.  Similarly, when users flag content as violating, it is sent to the content reviewers. Users have the opportunity to appeal the decision that is taken by the system. Enforcement of fake accounts is done in a separate workflow. }
\label{fig:enforcment}
\end{center}
\end{figure}

Figure~\ref{fig:enforcment} illustrates the flow of content on social networks and the enforcement of integrity. On the top, we have content and network creation, which includes new accounts, links between accounts (e.g., friendships), posts and reactions to posts. Ads are interleaved with posts on users' streams. Messages are exchanged between users via auxiliary applications in the network (e.g., Direct Messages on Twitter, Facebook Messenger). 

Potential violations of integrity are detected in two main ways: reports from users who see the violating content and AI systems that inspect the content as it streams onto the network.
Content that the AI system deems violating with very high confidence may be immediately removed.  When content is flagged, it may get demoted by the network in order to limit its virality while it is being verified. Demotion is done by down-ranking the content on users' stream so fewer people see it.  

Letting violating content stay on the network can have very severe consequences to individuals and to society. Hence, machine learning for integrity must err on the side of increasing recall and then use content reviewers to make the final decisions. 
Potential violations are checked in two avenues depending on whether they are community standard violations or misinformation that requires more expert knowledge. In the former case, the content is sent to a large pool of paid content reviewers who are trained to recognize the violations prohibited by the social network and to confirm which content should be removed. Content will be removed if multiple reviewers agree that it is violating. Factual claims require more expertise and are sent to third-party content reviewers. For example, in Facebook's case, these reviewers are certified by an independent body, the International Fact-Checking Network. They review and rate content via primary and secondary research to find evidence that corroborates or does not corroborate a statement of fact. If the machine learning model detects false content identical to content already rated by fact-checkers, it will apply fact-checks directly to the duplicate. This preserves the time of fact-checkers to focus on net new claims. 

The process described above is proactive--it handles content as it gets uploaded to the network or flagged by users. The system may also take more reactive steps, such as reconsidering certain posts after they have achieved high levels of engagement. The rationale for doing that is twofold. First, the engagement data may provide a strong signal about the validity of the post (e.g., a post having 15 disbelief or fact-checking comments at 1000 impressions is more likely to be a hoax). Second, content that reaches a wider audience on the network is more likely to cause harm if it is misleading or hateful. 

Removing content is the main mechanism that social networks use to enforce their policies, but it is also worth mentioning how social networks reduce the distribution of content. In particular, the social network demotes or downranks content lower in users' news feeds, to reduce its distribution and the likelihood that users on the platform will see it. Similarly, the social network can make it harder to re-share and forward certain pieces of content. This applies to content that is not eligible for removal, but is otherwise low quality for a user like clickbait, engagement bait and false news.

Another method for protecting users is based on the observation that some integrity violations start from search activities. In particular, inappropriate interactions with children and selling of illegal goods often begin with bad actors searching for vulnerable individuals or users searching for products for sale. Hence, some integrity problems can be partially mitigated by making certain searches harder to conduct.

\subsection{Measurement}
\label{sec:measurement} 

In order to assess the efficacy of the techniques for identifying violating content, we need a set of metrics we can track. Unlike many other machine learning applications, the adversarial nature of integrity and the fact that some violations happen with extremely low frequency, makes it tricky to design meaningful metrics. We discuss some of the measurements and their shortcomings below. Facebook publishes a report based on some of these metrics every six months~\cite{facebook-transparency-report}. 

\medskip
\noindent 
{\bf Prevalence:} prevalence, measured as a percentage of all content on the network, refers to the amount of content that is on the social network and was not caught by the enforcement mechanisms. The simplest way to measure prevalence is to count the number of distinct posts on the network. However, since some posts are viewed more than others, a more meaningful measure, {\em prevalence of bad experiences}, is the number of times violating posts have been seen by users. Experience prevalence can also be refined to take into consideration the severity of the violation. For example, a completely nude photo of a person would be considered a more severe violation than a photo that has only a partial view of a naked body. Of course, considering severity requires that there is  a method to attach a severity measure to each post for each type of violation. Another imperfection of experience prevalence is that it does not consider who the user is. Some users may be more negatively affected by certain violations than others. Prevalence, like recall of web documents, can be tricky to measure, so it is typically done with respect to some sample. 

\medskip
\noindent
{\bf Proactive rate: } what percent of the violating content was found by the algorithms before it was reported by users. In contrast to prevalence, this measures how much violating content the system finds, versus the amount that is missed. 

\medskip
\noindent
{\bf Auto-deletion volume: } how many violating posts were deleted by the algorithms automatically without human review. Auto-deletion requires that the confidence of the classifier be very high. This measure contributes to the goal of reducing the workload on human labeling.    

\medskip
\noindent 
{\bf Appeals:} what percent of content that was deemed violating was appealed and what percent of the appeals got reversed. 

These measures do not adequately address all needs. For example, several high severity but low prevalence problems like child safety are not measurable in terms of prevalence. Instead, a different measure that is used as a proxy is ``action rate'', namely, the percentage of actions taken that were enqueued via AI. 

When designing these metrics, it's important to keep in mind some of the unique aspects of the integrity problem. For example, on one day, the system may catch a set of violating content, notify its authors and remove the content. Two things may happen next. First, since many originators of violating content do not actually do it on purpose, these well meaning actors will not create violating content subsequently. Second, the bad actors who are trying to trick the system's defenses (sometimes for monetary benefit) may get more sophisticated and will change their strategy. The net result of these two behaviors is that the prevalence may increase in the following days. While a traditional machine learning setting may consider that a bad outcome, in the case of integrity, fluctuations in prevalence and other measures are to be expected.

\section{A holistic view of integrity}
\label{section:framework}

As described in the previous section, the problem of preserving integrity is extremely broad and covers many specific violations. These types of violations have been discovered over a long period of time as they arose in practice. In response, social-media companies have typically formed individual teams to handle each violation category in isolation. 

In this survey we take a holistic view of the landscape of integrity violations. Instead of discussing individual violation types, we identify key aspects of the social-media eco-system, each of which is common to a wide variety violation types. Furthermore, each of these components represents an area for research and development, and the innovations that are found can be applied widely.
In addition to providing clarity on the large body of work in this field, this view of the integrity landscape should also help in addressing violations we discover in the future.

\begin{figure}[ht]
\begin{center}
\includegraphics[width=5in]{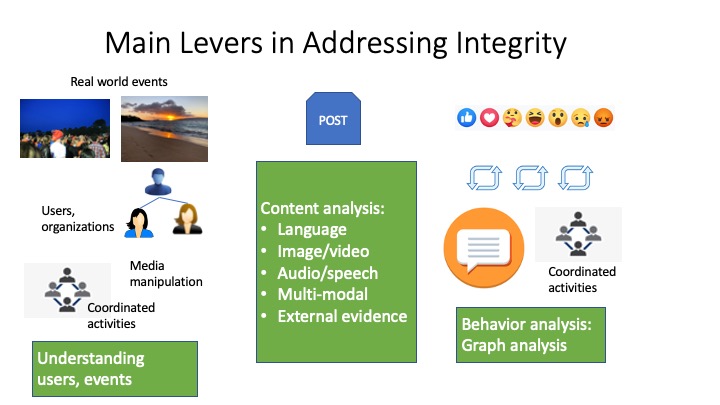}
\caption{In order to address a large and evolving number of integrity problems, it is important to identify the key technical levers that are common across problems. These levers can be broken down to ones that occur before the post is made (such as analyzing the underlying events happening in the world and possible manipulations made to the media before they're posted), signals that are available in the post itself (such as the text and the images), and the behavior of users in response to the post.   The behavior can either be {\em organic}, referring to the reactions of people who see the post, or {\em coordinated}, referring to groups of actors who purposely promote a post, or the behavior of the originator of the post in other contexts.}
\label{fig:framework}
\end{center}
\end{figure}

Figure~\ref{fig:framework} depicts our view of the integrity landscape, and distinguishes between the origins of the post, the post itself, and the reactions to the post.

\medskip
\noindent
{\bf Origins of the post:} There are several aspects concerning the origins of a post that are important in detecting violating content.
The first aspect is the real-world event on which the post may be based (e.g., a shooting, an election, a pandemic). The underlying event context provides a hint as to whether the post may be spreading misinformation, voicing hate speech after a terrorist attack, or trying to help a person who is in dire need. The second aspect is 
understanding the user or the organization making the post. For example, a user or group that has posted violating content in the past may be prone to do so more often in the future. Furthermore, that user may be part of a bigger community that is intentionally coordinating their actions to ensure that the content of the post gets wide distribution. Finally, media may be manipulated prior to posting to make it more compelling, and it is important to understand whether these manipulations are malicious.

\medskip
\noindent
{\bf Contents of the post:}
Naturally, much of the work on integrity focuses on the content of the post because the content contains the strongest signal. 
This body of work pushes the state of the art on the analysis of multiple modalities, including text, images, video and audio and the combinations thereof. Posts that contain multiple modalities (e.g., text and images) present unique challenges because each of the modalities may present benign content on their own, but it is the combination that reveals their true meaning. 
 We cover work on content analysis in Section~\ref{section:content}. 

\medskip
\noindent
{\bf Behavior after the content is posted:}
Integrity issues in social networks are usually facilitated by the social network itself by spreading and commenting on content. Hence, the behaviors of different actors on the network provide an important signal that increases the effectiveness of integrity violation classifiers.

We distinguish between {\em organic} behavior and {\em coordinated} behavior. Organic behavior refers to how users interact with a post after its published such as reacting (e.g., like, hate, sad), re-sharing a post (often with additional commentary), and conducting discussions by writing follow up comments. Understanding the nature of the actors involved in these reactions and the relationships between them can help distinguish between benign content and violating content.

Coordinated behaviors refer to sets of actors that together try to ensure that a post gets wider distribution so it appears more authoritative or simply reaches a much larger audience. For example, actors might band together to spread misinformation about an election in process  (e.g., wrongly stating that voting locations are already closed). As another example, pages might start out by posting benign content that is targeted at a particular user group. Once they have gained the trust and wide following of the group, they might start posting violating content targeting that group. We cover the study of behaviors and the understanding of actors on the network in Section~\ref{section:behavior}. 

Finally, there are several other issues that cut across many integrity problems, such as detecting violations while preserving privacy and supporting the human workforce of annotators. We cover these issues in Section~\ref{section:others}.

\section{Analyzing contents of posts}
\label{section:content}

Analyzing the content of a post is at the core of the algorithms for detecting violating content. Recognizing violations poses significant technical challenges to text understanding, image and video analysis. When multiple modalities are combined, as they are in memes, there are additional challenges. 

In this section we cover the main challenges and techniques in each of these. We begin in Section~\ref{subsection:text} with text analysis techniques. In section~\ref{section:nuances} we consider aspects of the text that are not explicit, such as the affect, style and subjectivity. Section~\ref{section:vision} discusses recent developments in computer vision that have made an impact on detecting integrity violations including the emerging area of manipulated media. 
Section~\ref{section:multi-modal} discusses the challenges that arise when combining multiple modalities. Finally, Section~\ref{section:external} considers how data that is external to a post (e.g., knowledge bases) can be used in identifying violations.

\subsection{Text Understanding}
\label{subsection:text}

Semantic understanding of text plays a key role in classifying whether a post is violating or not. Building on the long history of text classification research in the literature, many studies have been proposed to address different aspects in understanding text~\cite{lewis1990representation, joachims2001statistical, aggarwal2012survey, minaee2020deep}. Traditional approaches based on using bag of words models or a TF/IDF weighted embedding method can certainly help. However, text on social media and the nature of violations present several challenges to traditional approaches: (1) the text contains noise in the form of misspellings, orthographic variations and colloquial expressions (notably, the noise can be inserted by mistake or inserted on purpose in an attempt to confuse the detection systems), (2) policy violations are often intricate and classifying them requires deep understanding of the text, (3) labeled data is often limited, and (4)  content (violating or not) is created in many languages and creating per-language data set is infeasible.

Recent advances in self-supervised training for text~\cite{peters_deep_2018, devlin2018bert, yang_xlnet_2019, liu2019roberta} are showing great promise to address the issues outlined above. These techniques are more robust to text variations, they can capture context, and have shown some awareness of factual and common-sense knowledge \cite{petroni2019language, xiong2019pretrained,fbai-integrity-blog-19}. Self-supervised methods have achieved higher accuracy with the same amount of training data and can often outperform traditional methods when trained on less data. To complement these advances, new multilingual methods have shown that training data from one language can be utilized to learn classifiers that work on other languages not seen in the training data with increasing level of accuracy.

Self-supervised models for text work in two stages: 
\begin{itemize}
\item {\bf Pretraining stage.} Models are first trained on large quantities of unannotated text by hiding parts of the input and training the model to predict what is hidden. For example, in the traditional language model setting the model is given a prefix of a piece of text and is trained to predict the next word of the sequence (e.g., given ``{\em The cat sat on \_}'' predict: ``{\em the}''). The first successful self-training approaches for text such as ELMo~\cite{peters2018elmo} and GPT~\cite{radford2018gpt} use a language modeling objective. 

BERT~\cite{devlin2018bert} extended these techniques by introducing a masked language model objective. In this setting, a certain portion of the input text is selected at random and masked, and the model is asked to predict the masked while having access to the entire sequence  (e.g., ``{\em The cat \_ on \_ mat .}'' predict ``{\em sat, the}'').

\item {\bf Fine-tuning stage.} Following pretraining, the model is tuned to perform a particular task (e.g., detect hate speech) using labeled data. BERT does that by using the same model architecture as pretraining, but adjusting model weights to optimize classification accuracy (or related metric) on the training data.
\end{itemize}

BERT~\cite{devlin2018bert} has become the standard architecture for accurate text understanding. Its recent refinements, such as RoBERTa~\cite{liu2019roberta}, ALBERT~\cite{lan2020albert} and T5~\cite{raffel2019exploring}, have improved the training recipes and scaled to more data and parameters, thereby pushing the state of the art further. These approaches are especially helpful for difficult tasks like identifying hate speech because of the nuanced understanding of language that is required. 

The multilingual challenge has been addressed as the problem of Cross-Lingual Understanding. In this setting, a model is trained to perform a task using data in one or more languages, and is then asked to perform the task on data in other languages that are either not present or are underrepresented in the training data. Progress on cross-lingual understanding has been spurred by the introduction of several recent benchmarks for tasks, such as cross-lingual natural language inference~\cite{bowman2015large,williams2017broad,conneau2018xnli}, question answering~\cite{rajpurkar-etal-2016-squad,lewis2019mlqa}, and named entity recognition~\cite{Pires2019HowMI,wu2019beto}. 

Self-supervised methods have been successful at tackling the multilingual challenge starting with multilingual BERT (mBERT)~\cite{devlin2018bert}. mBert uses a single shared encoder to train a large amount of multilingual data. Further refinements, such as XLM~\cite{lample2019cross} and XLM-R~\cite{conneau2019unsupervised}, have closed the gap between in-language performance and performance on languages unseen during training data. XLM-R, in particular, has demonstrated that a multilingual model trained for 100 languages looses only a little in terms of accuracy (1.5\% on average) when compared to a model specialized for a particular language on a variety of tasks. Compared to mBERT, XLM-R was trained on an order of magnitude more training data and refined the training procedure by studying what training settings and hyperparameters are important and work well for multilingual learning. Theoretically, the success of multilingual models has been boosted by the fact that self-supervised methods learn similar spaces across languages (in terms of the topography of the underlying hidden embedding space used by the neural models)~\cite{wu2019emerging}. Cross-lingual models such as XLM have been successfully used for problems like hate speech classification across languages~\cite{fbai-integrity-hatespeech-blog-20}. 

\subsection{Nuances in text}
\label{section:nuances}

The work described in the previous section aims to understand the content of a piece of text. However, analyzing nuances in {\em how} the content is conveyed can provide an important signal about the intent of the author and the true meaning of the text. In this section we consider several such nuances.

\subsubsection*{Affect}
Affect in text (and otherwise) is classified along two dimensions: valence and arousal.  Valence refers to the inherent goodness of the topic discussed and can vary from being very negative to being very positive. Arousal refers to how activated the author sounds about the topic and can vary from  apathetic to being very excited. Certain emotions can be mapped to different quadrants of affect. For example, ecstatic would be classified with high arousal and valence, while depressed would be low on both dimensions. 

Intuitively, the emotions conveyed in a post can provide a signal as to its real intent. For example, if the language used in a post involves emotions with high arousal (e.g., anger), then that might indicate a more pronounced intention to hurt someone. Researchers are just beginning to consider affect for integrity problems. We describe two such research works below.

 Guo et al.~\cite{DBLP:journals/corr/abs-1903-01728} consider the value of detecting emotions in the context of misinformation. 
 They show that emotions should be considered in the content of the post itself and in the reactions (e.g., comments) to the post. In the case of high emotion in the post, the purpose of the publisher may be to arouse the audience. In cases that  the post is devoid of emotions, but there is disbelief in user comments, that may indicate that the post conveys misinformation. In their work, they train a classifier for sadness, happiness, doubt, anger and none. As they show, in the contents of the news item, anger is higher by 9\% in  fake news than regular news, and happiness is 8\% lower. The same trend appears in the comments to the article, and in addition, sadness and doubt are more pronounced. Based on these observations, they propose a system for predicting fake news based on content and on emotions. 
 
 Rajaramickam et al.\ consider the problem of adding affective features to the task of predicting abusive language~\cite{mishra20}. Their system is based on a multi-task learning approach~\cite{DBLP:journals/ml/Caruana97}, where the system is trained on a main task (in their case, abuse detction), but is also trained on secondary tasks (emotion detection). Their experimental results show that the joint learning task produces better results than a learning system trained on the single task of abuse detection, suggesting the system was able to learn affective features that are relevant to predicting abusive language.
 
\subsubsection*{Style}

The Undeutsch hypothesis~\cite{undeutsch1967beurteilung} suggests that the style of writing in misinformation potentially differs from the style of real news. This hypothesis has lead to a body of work in natural language processing that tries to distinguish styles of text in news articles. Researchers have developed supervised classification methods that use stylistic clues to identify when language may be deceptive~\cite{tausczik2010psychological}, and whether it is subjective or emotionally charged~\cite{zhou2019fake}.

In the case of identifying deceptive language, Mihalcea and Strapparava~\cite{mihalcea2009lie} analyzed the counts of certain words in text. They identified dominant word classes based on Linguistic Inquiry and Word Count (LIWC)~\cite{pennebaker2001linguistic} in deceptive or truthful text. For instance, they observe that the truthful text is more likely to contain words from the `optimistic` word class (e.g., hope, accept, determined) whereas the deceptive text is more likely to contain words from the `certain` class (e.g., always, all, very, truly, completely, totally). Ott {\em et al.}~\cite{ott2011finding} extend this idea by considering linguistic features such as part-of-speech tags present in the text and psycho-linguistic cues (e.g., the average number of words per sentence, the rate of misspelling, and references to terms like work, money, religion).  Their classifiers, which are trained on features traditionally employed in (a) psychological studies of deception and (b) genre identification (such as imaginative writing), outperform ngram–based classifiers.

Similar methods have been applied to detect emotive and subjective language for the purpose of identifying fake news. Detecting subjectivity in text has been studied extensively in the area of opinion mining and sentiment analysis~\cite{wilson2005recognizing, lin2011sentence}. The work of Libanio {\em et al.}~\cite{jeronimo2019fake} hypothesizes that subjectivity levels will differ significantly between legitimate and fake news. They build subjectivity vectors for each news article using the Word Mover's Distance (WMD)~\cite{kusner2015word}. The WMD distance metric computes the minimum distance that a word from a document needs to ``travel'' to reach a word in another document in the embedding space. To build the subjectivity vectors, they first identify 5 subjectivity categories (argumentation, presupposition, sentiment, valuation and modalization), and associate a set of words with each category. They consider each category to be a document and find the WMD distance of a news article to each of these documents to build a 5 dimensional vector for the news article. These subjectivity vectors are finally used to train classification models. Their results show that their approach outperforms n-gram based classifiers, especially when training and test sets are from different domains.

Potthast {\em et al.}~\cite{potthast2017stylometric} analyze the writing style of hyper-partisan (i.e., extremely one-sided) news in connection to fake news. They use a manually curated dataset of articles that were fact-checked by journalists. The corpus contains 299 fake news articles, 97\% of which originated from hyper-partisan publishers. They show that the style of left-wing and right-wing news have a lot more in common than any of the two have with the mainstream. They also show that hyper-partisan news can be discriminated well by its style from the mainstream. Even though their approach does not perform well for the task of detecting fake news, they suggest that identifying hyper-partisanship can be used as a pre-screening for fake news detectors.

\subsubsection*{Clickbaits}

The majority of the online news media generate revenue from the clicks made by their readers. To encourage readers to click on an article and subsequently visit the media site, some media outlets resort to using catchy headlines, known as clickbaits, to accompany the links to the article.  Clickbaits may involve deception, which can be an indicator of misinformation~\cite{chen2015misleading}. 

Understanding the characteristics of clickbaits has been of interest for linguists. For example, Blom and Hansen~\cite{blom2015click} study the linguistic phenomena used in headlines as a means to arouse curiosity. Their analysis of 2000 random headlines from a Danish news website identifies two common forms of forward-references: discourse deixis and cataphora. Discourse deixis are references at discourse level ``This news will blow your mind''. Cataphora, on the other hand, is at phrase level, e.g., ``This name is hilarious''. Based on a dictionary of basic deictic and cataphoric expressions, their study shows that these expressions occur mostly in commercial, ad-funded, and tabloid news websites. 

Given the above linguistic observations, several research works have trained classifiers that are based on the lexical and semantic features of clickbait headlines or bodies. Potthast {\em et al.}~\cite{potthast2016clickbait} define lexical and semantic features based on the teaser message for a corpus of annotated tweets. These features include bag of words, the sentiment and readability scores and dictionary based features. Dictionaries include a list of easy words and clickbait specific phrases and patterns. They also define features from the linked web page, and meta information. Instead of hand engineering features, more recent work apply different neural network models to the task of classifying clickbaits~\cite{anand2017we, thomas2017clickbait, zhou2017clickbait, kumar2018identifying}. Finally, Bourgonje {\em et al.}~\cite{bourgonje2017clickbait} take a very different approach and suggest that clickbait can be identified by determining the stance relation (i.e., ``unrelated'', ``agree'', ``disagree'', ``discuss'') of an article headline to its corresponding body. This proposal is based on the assumption that the body of a clickbait may not always be faithful to its headline.

\medskip
\noindent
{\bf Outlook:} In addition to the nuances described in this section, we believe there are quite a few aspects of text that can be useful for detecting violations. As a couple of examples, it can be informative to know whether the text in a post is making a reference to a famous or recent quote,  and whether the text is name calling (e.g., ``Republican congressweasels'' or ``caravandals'').

\subsection{Computer vision}
\label{section:vision}

\begin{figure}[!t]
\centering
\includegraphics[width=0.4\textwidth]{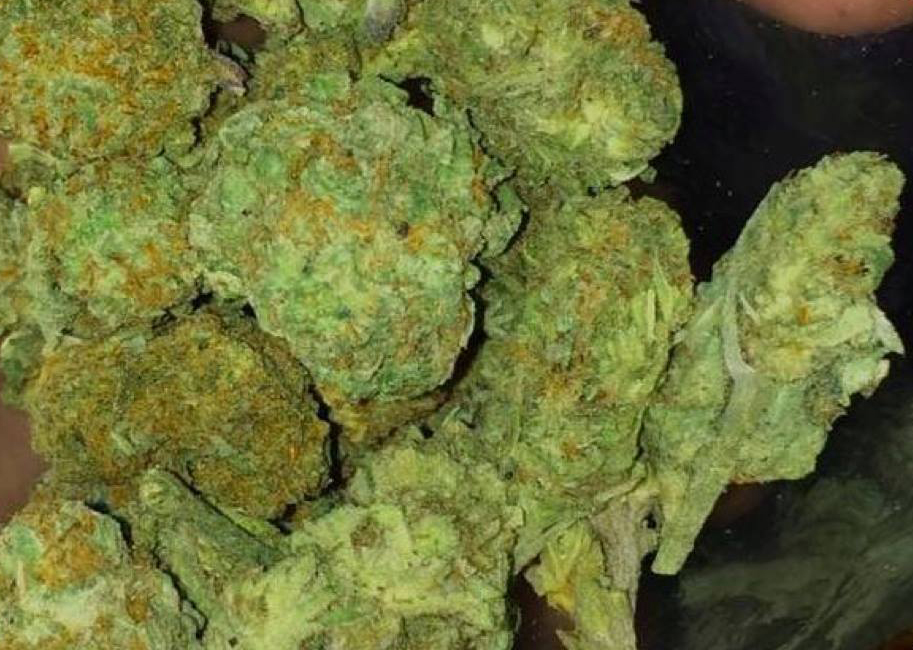} \hspace{1cm}
\includegraphics[width=0.4\textwidth]{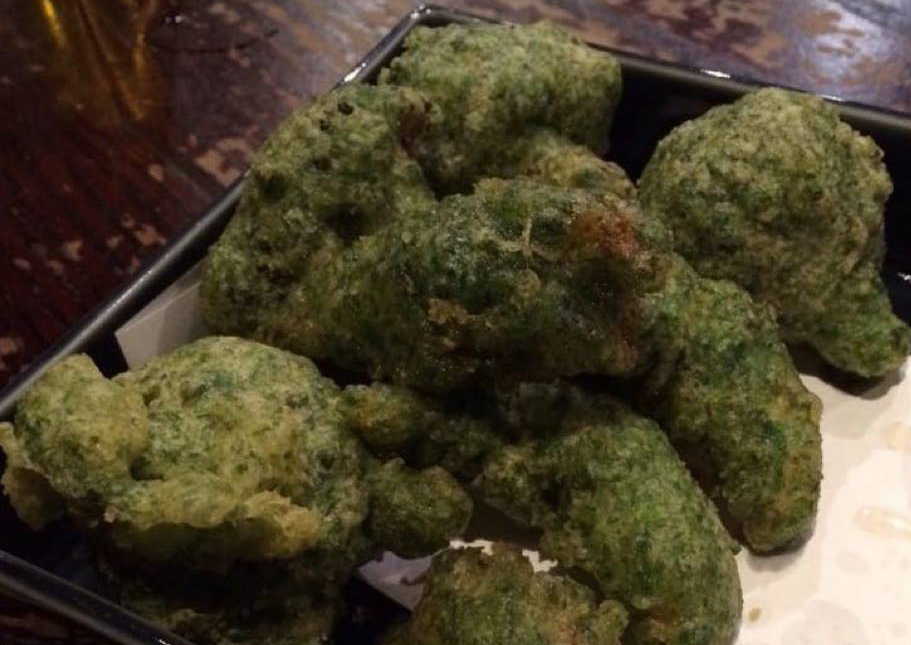}

\caption{Visual similarity between benign and violating content. One of the images above is of marijuana and the other is of fried broccoli. The similarity between the images highlights the need for high accuracy computer vision systems to address automatic detection of policy violations. The reader is encouraged to try to decide for themselves which of the images is the violating one. The correct answer is in Section~\ref{section:others}.}
\label{fig:marijuana}
\end{figure}

People share large amounts of photos and videos on social media and the Internet in general. Hence, understanding the content and context of what is represented in them plays a critical role to keep people safe. In particular, understanding the semantic content of images and videos at scale is critical for detecting integrity violations that involve objectionable content (e.g., nudity, pornography, graphic violence, gore) and regulated goods (e.g., drugs, weapons). 

Advances in computer vision have pushed the state-of-the-art in supervised learning to a point where it is feasible to do image and video understanding with a high degree of accuracy. Figure~\ref{fig:marijuana} illustrates an example where current computer vision systems can distinguish between benign content (fried broccoli) and violating content (marijuana), where it would be arguably tricky for humans to do the same.  

\begin{figure}[!t]
\centering
    \begin{minipage}[b]{0.45\textwidth}
    \centering
      \includegraphics[width=0.8\textwidth]{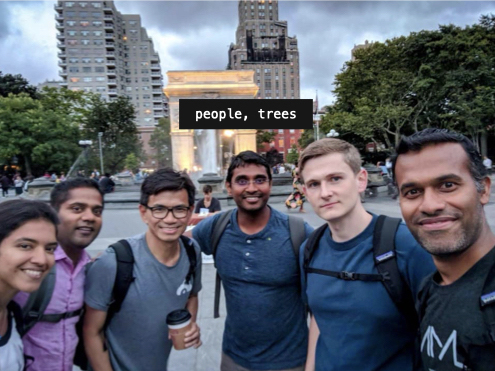}
      \par AlexNet
        
    \end{minipage}
    \hfill
    \begin{minipage}[b]{0.45\textwidth}
    \centering
    \includegraphics[width=0.8\textwidth]{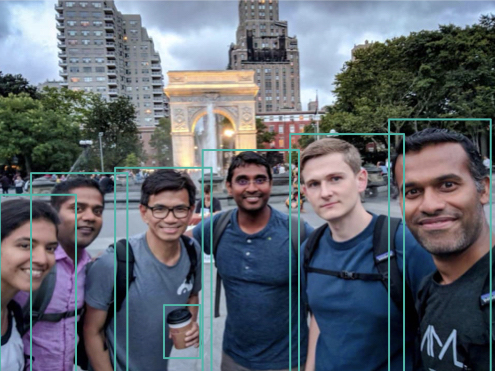}
    \par Faster R-CNN
        
  \end{minipage}
  
  \vspace*{1mm}
  \begin{minipage}[b]{0.45\textwidth}
  \centering
      \includegraphics[width=0.8\textwidth]{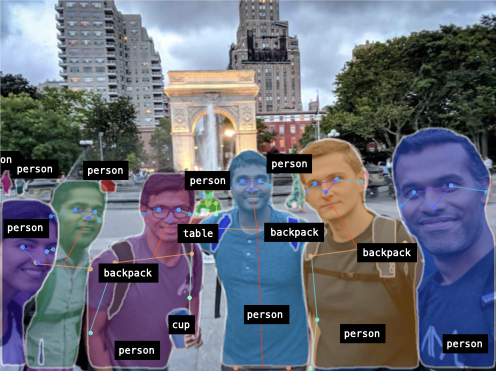}
         \par Mask R-CNN 
    \end{minipage}
    \hfill
    \begin{minipage}[b]{0.45\textwidth}
    \centering
    \includegraphics[width=0.8\textwidth]{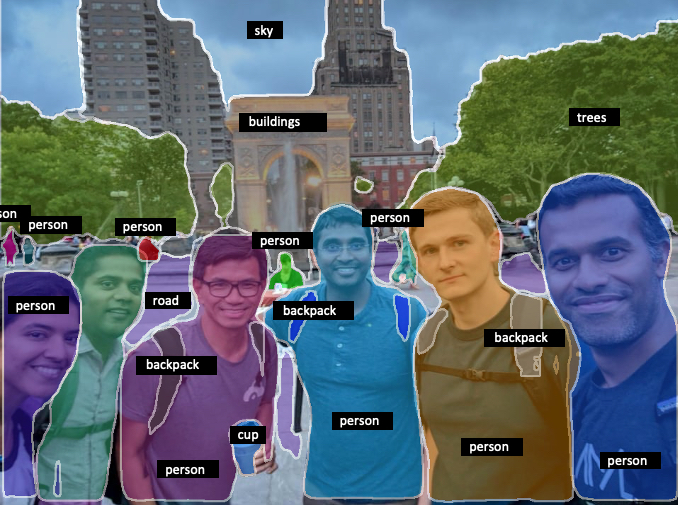}
    \par Panoptic FPN
    
  \end{minipage}

\caption{Different scene understanding methods (in order of complexity) used within computer vision. Parsing and recognizing different elements in images and videos allow detection of objectionable content and other policy violations.}
\label{fig:vision-techniques}
\end{figure}

In image analysis, as illustrated in Figure~\ref{fig:vision-techniques}, there are techniques for understanding images at different levels of granularity: the image as a whole~\cite{mahajan2018exploring}, where a standard CNN approach to classification is applied; coarse regions (e.g., people, objects)~\cite{he2017mask}, based on having a CNN that proposes candidate object regions (rectangles) in the image and another CNN that classifies those regions; and at a per-pixel level \cite{kirillov2019panoptic}, where a pyramidal refinement is applied to produce predictions everywhere in the image.

Analyzing video is a more challenging task since it requires understanding the semantics not only spatially, but also temporally and to consider audio when available. Often, videos are analyzed as 3D volumes (image+time) yielding compelling results~\cite{Tran_2019_ICCV}. However, the computational cost of analyzing 3D volumes has proven to be prohibitive and therefore researchers have developed techniques that factorize them into separate 2D convolutions (for images) and 1D convolutions (for time), while still achieving state-of-the-art accuracy~\cite{Tran_2019_ICCV}.

To understand the semantics of a video, a common technique is to split it into short clips and process them through a spatio-temporal understanding model. Information from multiple clips is aggregated in a later stage to produce a final semantic prediction for the video under analysis. However, a challenge arises from the fact that a given policy violation (e.g. bullying or child exploitation imagery) is typically found only in very few sections/clips of the video while the rest may feature benign content. Saliency mechanisms~\cite{korbar2019scsampler} have been found to be an efficient solution to surface policy violating videos even in cases where the violations do not dominate the entire video.

Research on objectionable content is scarce and mostly centered on the detection of nudity by skin detection~\cite{ap2005algorithm}, genitalia~\cite{10.1007/978-3-030-39575-9_31}, motion patterns~\cite{moreira2016pornography}, or expressive audio. Detection of violence is based on locating gory elements, while other more subtle cues like motion~\cite{10.1007/978-3-030-11012-3_24} have been recently proposed.

Child Exploitation Imagery (CEI) (a.k.a.\ child sexual abuse material (CSAM)) is one of the worst problems faced by integrity enforcement. Borrowing a technique from the field of privacy, robust hashing like PhotoDNA~\cite{photodna} has become the standard to efficiently locate child exploitation imagery. With PhotoDNA, we generate a set of unique hashes for a set of images (essentially, a block-list of images). This list of hashes is then circulated among many partners in the industry, who can then check for violating content on their platforms. Homomorphic encryption~\cite{singh2019robust} has been proposed as a viable solution where PhotoDNA computations are performed on-device without the further need of sharing the original image to the platform taking actions. Other solutions that have been applied to the problem include sensitive data storage and sharing~\cite{GuoDolhansky2020}, where violating media is represented in embeddings that have the property of being non-reversible, i.e., the original media can't be reconstructed from such embeddings. 

The scarcity of the research on objectionable visual content is partially due to the sensitivity of the data at stake and the difficulties of creating standard datasets. Only a few models have been publicly disclosed~\cite{yahoo-nsfw, basilio2011explicit} and this is still an area for further development. 

\subsubsection*{Manipulated media}
\label{section:manipulated}

Images, audio and video shared across social media are rarely posted without some level of editing. Minor edits to enhance the quality of the media (e.g. contrast, color enhancement, video stabilization) are often applied by default while other more evident manipulations are also pretty prevalent: beautification filters \cite{Leyvand2006, Liu2019FaceBB}, automatic makeup \cite{Guo2009, Chen2019}, or the addition of digital augmented reality elements into the scene. 

\begin{figure}[!t]
\begin{center}
    \includegraphics[width=0.2\textwidth]{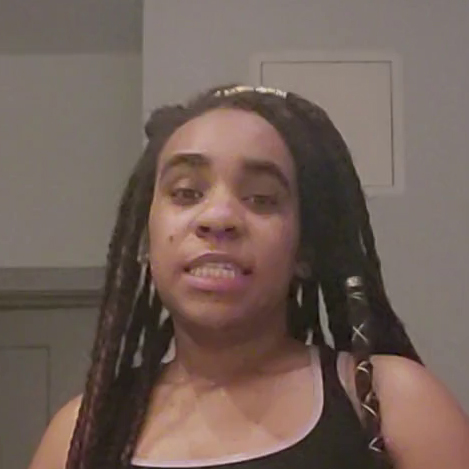}\includegraphics[width=0.2\textwidth]{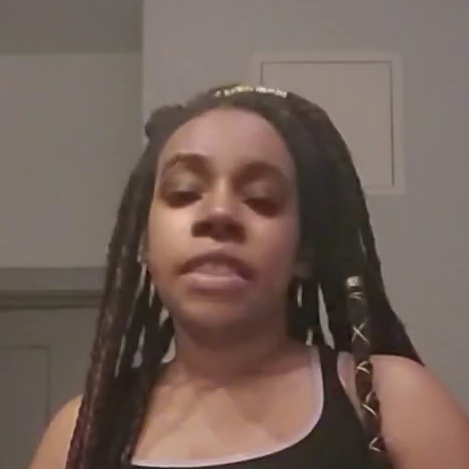}\hspace{1mm}\includegraphics[width=0.2\textwidth]{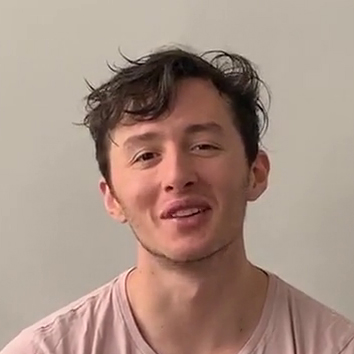}\includegraphics[width=0.2\textwidth]{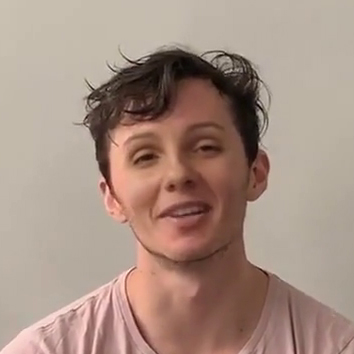}\\
    \vspace*{1mm}
    \includegraphics[width=0.2\textwidth]{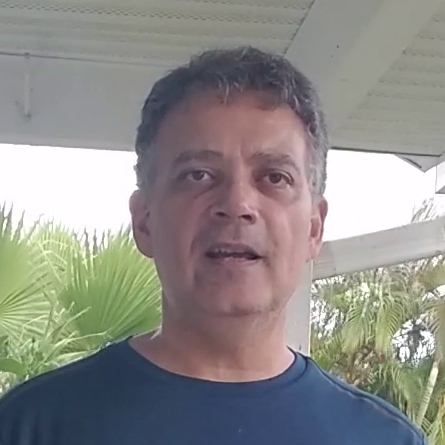}\includegraphics[width=0.2\textwidth]{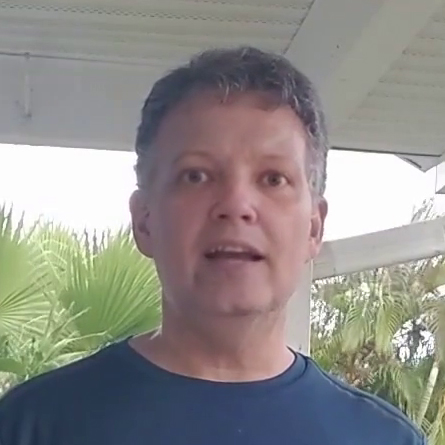}\hspace{1mm}\includegraphics[width=0.2\textwidth]{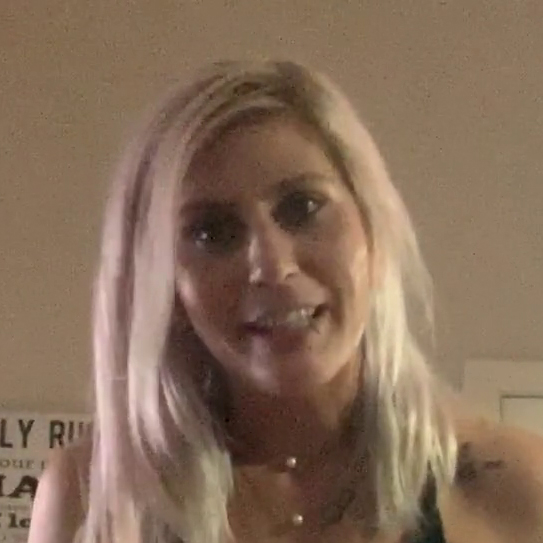}\includegraphics[width=0.2\textwidth]{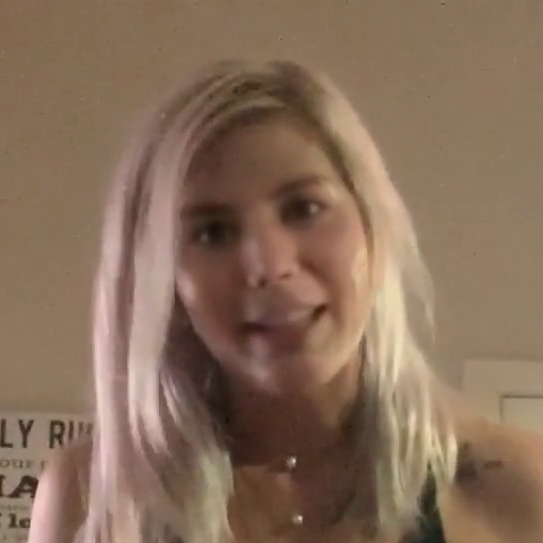}\\
\end{center}
\caption{Some DeepFake examples from the DFDC Lite dataset \cite{Dolhansky2019TheDD}. Left is original, right is the face swap.}
\vspace*{-2mm}
\end{figure}

With the recent progress in computer vision and speech processing and the advent of generative adversarial networks, manipulated media has leaped a significant step forward. These advances have unleashed potential for harmful applications like face puppeteering~\cite{Zeng2020RealisticFR, Nirkin_2019_ICCV}, speech manipulation~\cite{Polyak2019TTSSS}, face transfer~\cite{thies2019neural}, full body manipulation~\cite{chan2019dance}, and other more general media manipulations. Many of these manipulations can be used to impersonate others, spread false information, or just introduce bias in the observer. The newly coined term \emph{DeepFake} has been coined to describe these type of manipulations. 

Accordingly, research on detecting media manipulation has recently become a very prominent topic~\cite{DFDC2020, Verdoliva2020MediaFA}. To date, two main techniques have emerged in this field. The first is based on learning the traces that media modification methods leave in the resulting footage~\cite{guera2018deepfake, li2018exposing} of the media. Different generation methods introduce subtle artifacts in the media derived from interpolation, inaccuracies of the generation process or the post-processing, usually hardly perceivable to the naked eye. Although these methods evolve and improve every year, such subtle artifacts are still one of main ways to attack detection of deepfakes. The second set of methods is based on analyzing physiological cues associated with human faces. In these cases, manipulated or full synthetic images lack some subtle cues like blood flow~\cite{oh2018learning} or eye blinking~\cite{li2018ictu}, which can be detected by computer vision methods. Within this same field, motion cues associated with target individuals~\cite{agarwal2019protecting} are very distinctive and can be used as a soft biometric to assess the identity of someone, e.g., how someone smiles or talks.
 
\subsection{Multi-modal reasoning}
\label{section:multi-modal}

\begin{figure}[htb]
\begin{center}
\includegraphics[height=4cm]{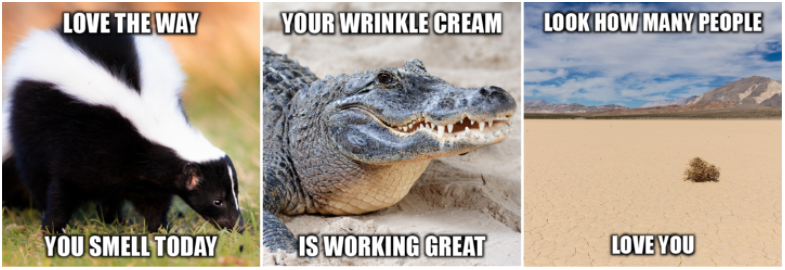}
\caption{In each of the three memes above, both the text and the images, taken alone, are benign, but the combination results in an ill-intended meme. }
\label{fig:memes}
\end{center}
\end{figure}

Detecting many types of violations, such as misinformation or hate speech, is often subtle because it relies on combining multiple pieces of information and possibly knowledge about the world. This issue is most pronounced in multi-modal posts and ads that combine text, images and videos, where the {\em combination} of modalities is what provides the real meaning of the content. Memes, narrowly defined as images with overlaid text, are a common form of multi-modal content that is designed to spread from person to person via social networks, often for (perceived) humorous purposes. As we illustrate with the memes in Figure~\ref{fig:memes}, it is often the case that the text and the image in isolation are benign, but when they are combined, their meaning changes and the content can become  objectionable. For humans, understanding memes is easy, but for a machine it becomes harder than understanding each of the modalities alone. Unfortunately, in practice, many hateful posts are based on memes.

The current state of AI technology in the area of understanding multi-modal content effectively or efficiently is still in its infancy compared to our ability to understand each of the individual modalities. Hence, the field of integrity is an important impetus to pushing the state of the art on multi-modal reasoning. The Hateful Memes Challenge Dataset~\cite{kiela-memes} has been created to foster research on the topic. 

\begin{figure}
\centering

\begin{minipage}[b]{0.45\textwidth}
    
    \includegraphics[width=\textwidth]{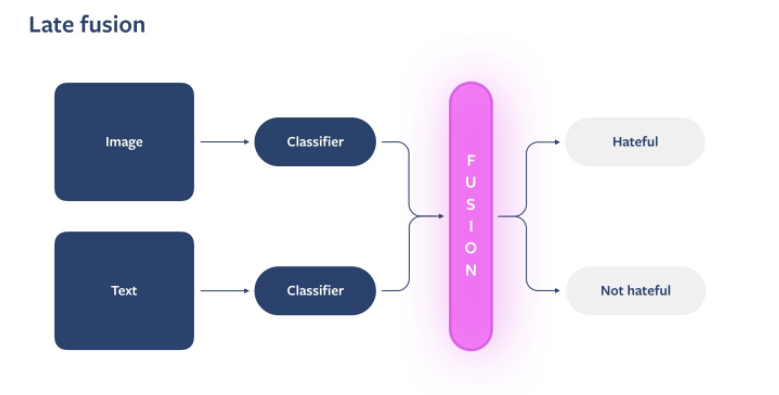}
  \end{minipage}
  \hfill
  \begin{minipage}[b]{0.45\textwidth}
    
    \includegraphics[width=\textwidth]{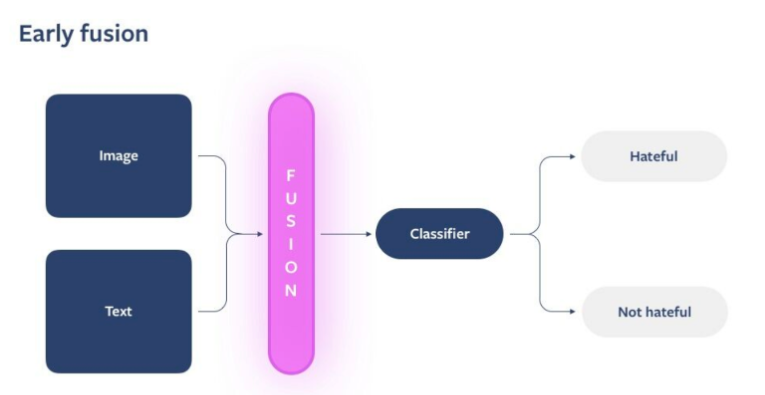}
  \end{minipage}

    \caption{A late-fusion classifier (on the left) uses existing uni-modal classifiers to make predictions and then combines the predictions in a fusion layer. An early-fusion classifier (on the right) feeds the raw data into a classifier and only then makes a prediction. While late-fusion systems are easier to build, they are ineffective at understanding content that combines multiple modalities in subtle ways.}
\label{fig:fusion}
\end{figure}

Recent research on the topic has started favoring classifiers based on early fusion over those based on late fusion (see Figure~\ref{fig:fusion}). A late-fusion classifier uses existing uni-modal classifiers and fuses them at the last layer. While they are simpler to build, they are ineffective at understanding content that combines multiple modalities in subtle ways. In contrast, early fusion classifiers feed the raw data into a fusion classifier before any predictions are made. 

One of the challenges in training classifiers that consider multiple modalities is that 
they are prone to overfitting to one of the modalities. One reason for overfitting is that one modality may dominate the content. For example, there are scenarios in which the textual part of the content dominates, i.e., posts don't always have an image/video attached but they almost always have text. In such a scenario, a naive model may overfit to the text, and as a result, will fail to take advantage of the corresponding image content to understand the full context. A second reason is that the optimal learning speed may be different for each modality, and therefore applying the same rate to all modalities will cause some to overfit. Gradient blending~\cite{wang20multimodal} has recently been introduced address this issue. The technique computes an optimal blend of modalities based on their overfitting behavior.

Advances in representation learning also offer benefits to multi-modal reasoning. The  Multi-Modal bitransformer~\cite{DBLP:journals/corr/abs-1909-02950} uses pre-trained unimodal representations (BERT and ResNet-152) and then fine tunes them together for the task at hand. VilBERT (short for Vision-and-Language BERT)~\cite{DBLP:conf/nips/LuBPL19}, on the other hand, pre-trains the model using both text and image data, extending self-supervised methods so the system can learn early how text refers to parts of an image and vice versa. The multimodal bitransformer advances the more nuanced understanding at the category-level such as flagging entire classes of content about drugs or other harmful content. In contrast, VilBERT pushes the accuracy of multimodal understanding of object-specific tasks like question-answering.

Several of the concepts mentioned above form the basis for the Whole Post Integrity Embeddings  System (WPIE) that is used in production at Facebook~\cite{blog-with-wpie}. 
To get a more holistic understanding of the post, WPIE is pretrained across multiple modalities, multiple integrity violations and over time. In a sense, just as cross-lingual pretraining can improve a classifier's overall performance, WPIE learns across dozens of violation types in order to develop a much deeper understanding of content. The system improves performance across modalities by using focal loss, which prevents easy-to-classify examples from overwhelming the detector during training, along with gradient blending, which computes an optimal blend of modalities based on their overfitting behavior. 

One of the unexplored challenges in multi-modal reasoning is that often an external piece of knowledge is needed in order to make the link between the different modalities. For example, consider the left image in Figure~\ref{fig:memes}. The crux of the meme is based on the fact that skunks are known to produce spray that stinks. However, the smell of skunks is not evident anywhere in the image, and because of that it may be hard to train a classifier to go through the complex reasoning process that combines the meaning of the text and the image. This example and others like it raise several challenges: (1) finding external knowledge sources that can complement the multi-modal model, (2) selecting the relevant piece of external knowledge (or small superset thereof) that should be added to the model, and (3) designing an architecture that can combine the new information in order to reach the desired conclusion. These are all  interesting research challenges that are ripe for exploration. The following section discusses cases in which external knowledge was used successfully to corroborate information in a post.  

\subsection{External evidence}
\label{section:external}

So far we discussed techniques for analyzing the content of a social-media post itself in order to determine weather it violates the integrity policies.  A complementary technique is to corroborate the content in the post with external information. Corroboration is particularly useful in detecting misinformation, where the language may sound credible, but the claims will contradict well established knowledge~\cite{thorne2018automated}.

In the context of misinformation, techniques relying on external information, known as evidence-based approaches, try to compare the claims made in the post with the existing evidence with known veracity. While some studies begin by extacting claims that are worth checking from the post~\cite{hassan2015detecting, hassan2017toward, konstantinovskiy2018towards, jaradat2018claimrank}, most work assumes that claims are given as input and focus on determining the truthfulness of those claims.

There are two main lines of work in evidence based verification of posts. They differ on whether the evidence being used comes from a structured knowledge base or is unstructured text.  In the structured representation, evidence is represented, as in traditional knowledge graphs, as a triple of the form (Subject, Predicate, Object), e.g., (Paris, capital\_of, France). We can then use large knowledge bases such as Freebase~\cite{bollacker2008freebase} as sources of ground truth because they contain knowledge that has been verified by humans. 
Works such as~\cite{vlachos2015identification, thorne2017extensible} study the identification and verification of simple numerical claims about countries (e.g., population or inflation) that are found in posts against Freebase. 

While appealing, there are several shortcomings in relying on structured knowledge bases for verification.  First, knowledge bases are often far from complete and they may not contain timely information about recent events. Second, converting claims made in text or speech into triples remains a very challenging NLP problem. Third, while some of these claims could be represented as single triples, real-life claims typically require more complex representations. For example, consider representing the claim ``Hillary Clinton flew to Moscow to discuss cooperation with Putin just before the Democratic National Convention'' as a set of triples. Even if it was possible do translate a claim reliably into a set of triples, they may not all be of equal importance (e.g., what if they met in Istanbul instead of Moscow, would it change the spirit of the claim?).

In an unstructured representation, evidence is assumed to be textual claims. Here the evidence can come from encyclopedia articles, policy documents, verified news (e.g., articles from PolitiFact\footnote{www.politifact.com} or FactCheck\footnote{www.factcheck.org}) and scientific journals. Vlachos and Riedel~\cite{vlachos2014fact} proposed to frame the verification task as a sentence similarity task between existing claims and the claims that are being verified. However, they did not specify the exact similarity labels needed for this task. Ferreira and Vlachos~\cite{ferreira2016emergent} use article headlines as evidence to predict whether an article is supporting, refuting or observing a claim. The Fake News Challenge~\cite{pomerleau2017fake} also uses textual evidence for fact checking the claims, but the entire documents are used as evidence. This allows for evidence from multiple sentences to be combined to compare with a claim made in an article. 

The above works rely on verified evidence and are therefore restricted only to claims that are  similar to the ones already available in the dataset. The Fact Extraction and VERification (FEVER) task~\cite{thorne2018fever} goes further by requiring the system to combine evidence from multiple documents and sentences. In this task, the evidence is not given and needs to be retrieved from Wikipedia.

Going beyond misinformation and the verification of claims, we believe that an interesting avenue for research is to endow integrity algorithms with broader knowledge about the real world and its subtleties. In particular, deciding whether a post is violating often requires knowledge about current events, long-standing conflicts and troubled relationships in the world and sensitivities of certain populations. While creating such a knowledge base about the world is certainly imaginable (and would be useful even if it is not complete), there would be several challenges to incorporating such knowledge into violation detection. For example, we would not want to assume that just because someone is a member of a particular population, they are sensitive to the same content. Furthermore, while general knowledge may shed light on a specific situation, the details of the particular post and time are critical.  

\section{Analyzing network behaviors}
\label{section:behavior}

The analysis of network features plays a key role in recognizing violating content. In a sense, this should come as no surprise since the network is the medium that is used to disseminate and amplify  the content and sometimes to modify the intent of the original post. 

There are two aspects to analyzing network effects. The first is understanding user interactions with a post after it is published. 
Users interact with content through several mechanisms: reaction emojis (e.g., like, haha, angry, sad), commenting on a post (which, in turn, can lead to conversations and arguments), and re-sharing a post with their own network. As a result, content (violating or not) generates  some reaction in the real world, and the nature of this reaction provides an important signal as to whether the content is violating. 

The analysis of the network reaction is particularly useful in the cases where the content itself is inherently ambiguous. In such cases, it may not be possible to classify a post as violating at the time of posting. However, once some user reaction has been generated and analyzed, violating content can be removed to reduce further damage. Of course, even if the violation was not caught in time, post-hoc analysis provides additional training data for future cases. We discuss the analysis of user reactions in Section~\ref{section:reactions}.

The second type of insights that a network-level analysis can reveal is an understanding of the actors on the network (i.e., users, groups, organizations) and relationships between them. For example, some actors may have been involved in violations in other places and times on the network, or may belong to communities prone to creating violating content. In extreme cases, some actors may be part of coordinated attempts to manipulate the visibility of some posts. We cover these kinds of analyses in Section~\ref{section:actors}. 

Many of the techniques for analyzing network behaviors are based on graph representations of different attributes of the social network, and we will discuss them in more detail throughout the section. 
Figure~\ref{fig:socialnetwork} depicts some of the entities and the relationships that are commonly used in these techniques. 

\begin{figure}[ht]
\begin{center}

\includegraphics[width=5in]{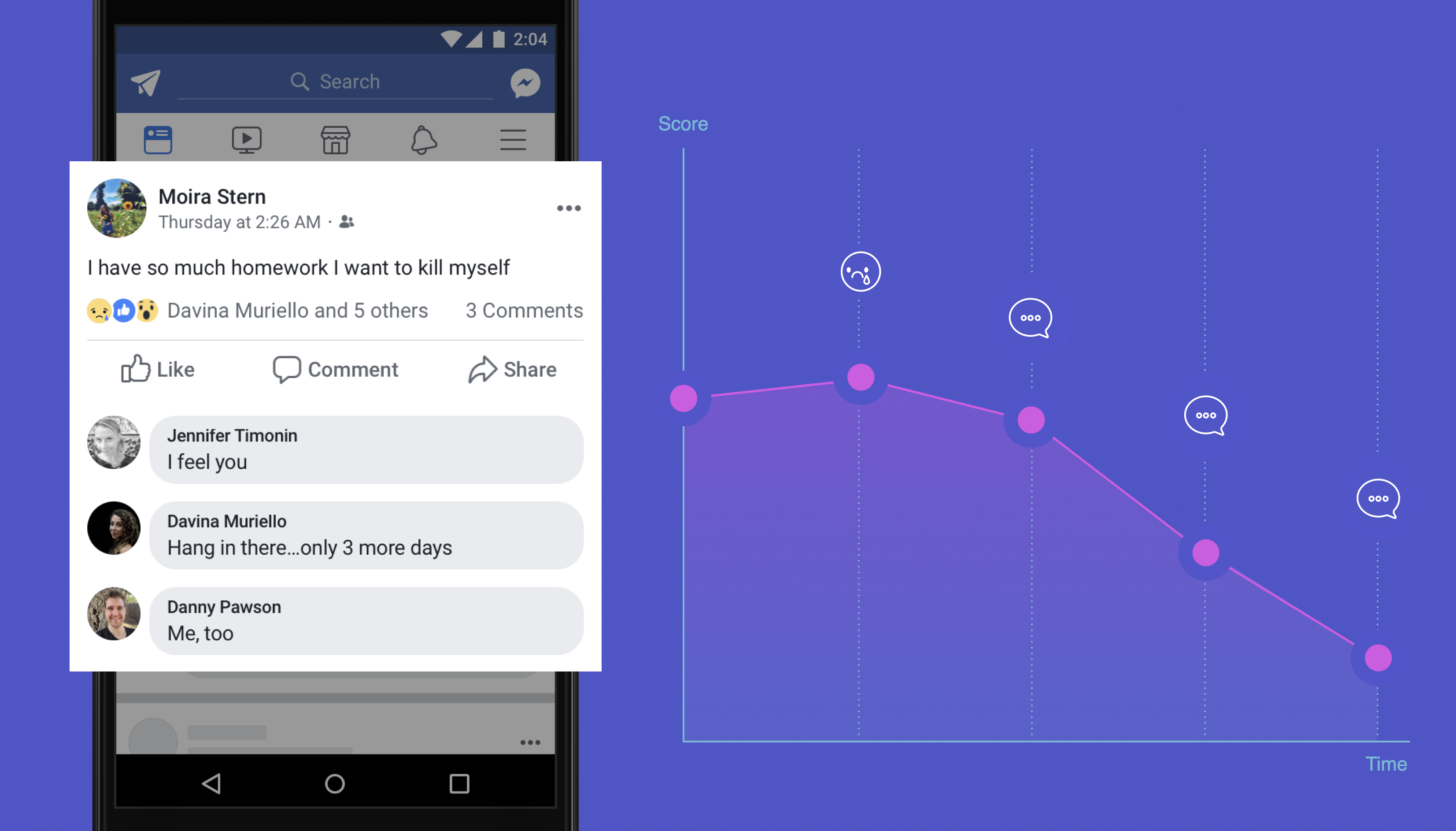}

\caption{The user mentions the desire to commit suicide, but the reactions from her friends make it clear that she used the term as a figure of speech, and hence the system's predictions for whether this is a suicide attempt decrease as the comments roll in.}
\label{fig:suicide}
\end{center}
\end{figure}

\begin{figure}[ht]
\begin{center}

\includegraphics[width=5in]{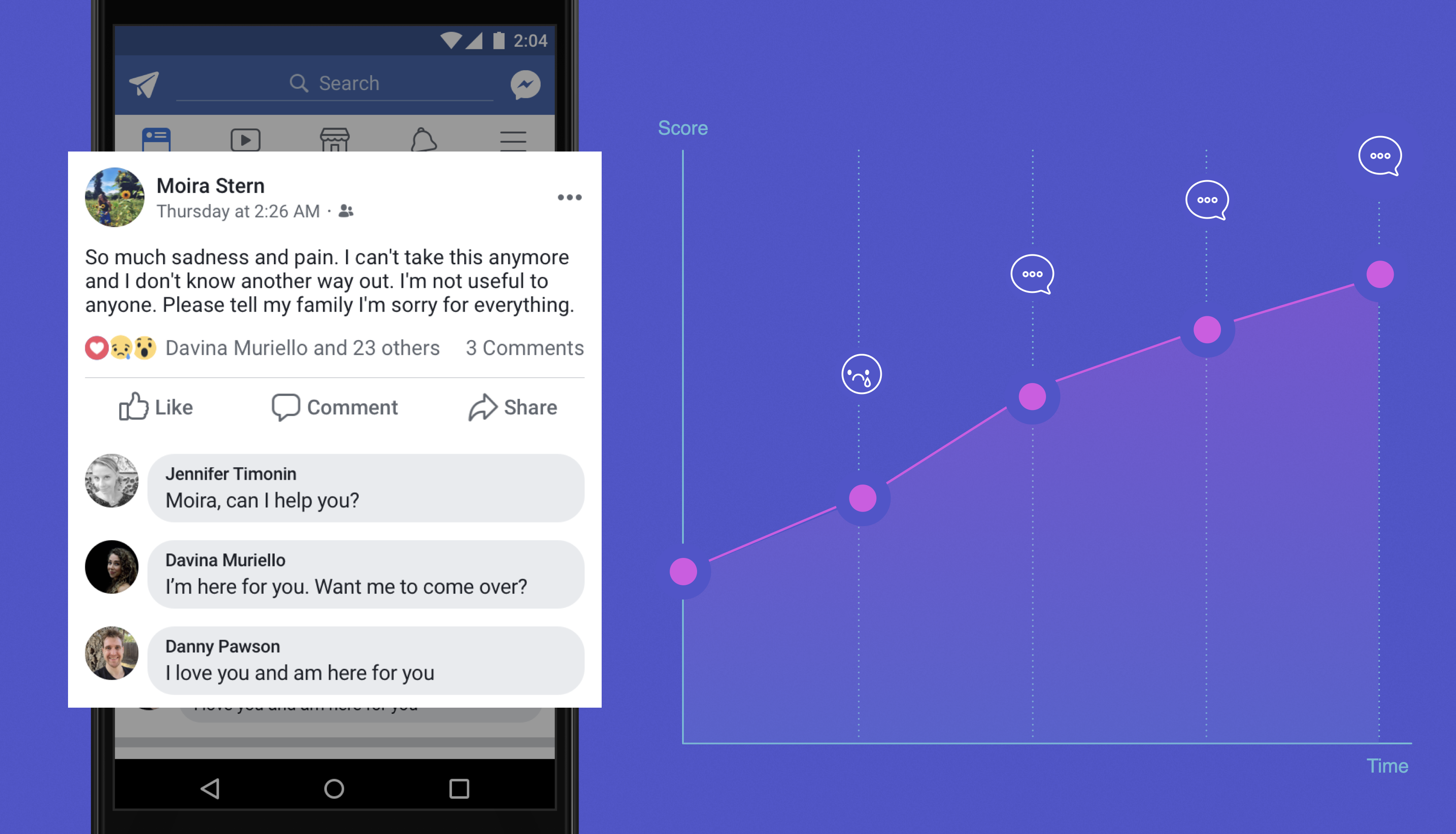}
\caption{The user is in pain but only makes subtle references to an attempted suicide. However, the reactions from her friends make it clear that the situation could be grave. As a result,  the system's predictions for whether this is a suicide attempt increase as the comments roll in.}
\label{fig:suicide2}
\end{center}
\end{figure}

\subsection{Reactions to posts}
\label{section:reactions}

Reactions to posts are extremely useful for classifying their intent because many posts can be ambiguous. For example, misinformation can often look like valid news articles, but the disbelief (or debunking) comments by users can serve as an important counter signal. Similarly, some posts may contain seemingly benign jokes about certain groups of people, but the comments may be racist or sexist, e.g., ``haha soo funny. we should deport all these rats'', further bringing into question the intention of the original post.

A particularly clear example of the value of analyzing reactions is in the case of detecting suicide attempts by users. The prevalence of these incidents is relatively small which makes it hard to gather enough training data for classifiers. However, the severity of the consequences of such posts is extremely high. To complicate things further, the  language of a suicidal post can be very subtle. On the other hand, reactions from friends and relatives can be very telling. When someone is concerned about their friend or family member who makes a post, they are a lot more direct and explicit (e.g., ``Where are you right now?'', ``I am trying to reach him and his phone is not answering'', ``we should call the police'', ``we should call his mom'').  Figures~\ref{fig:suicide} and~\ref{fig:suicide2} show example scenarios where the interactions are shown on the left, and the change of the predicted score for possible suicide is shown on the right. In Figure~\ref{fig:suicide}, the poster explicitly talks about suicide, but as it becomes clear from the reactions, it was used as a figure of speech. In Figure~\ref{fig:suicide2}, the language is much more subtle, but the reactions of the user's friends make it clear that the situation could be grave.

\begin{figure}[ht]
\begin{center}
\includegraphics[width=5in]{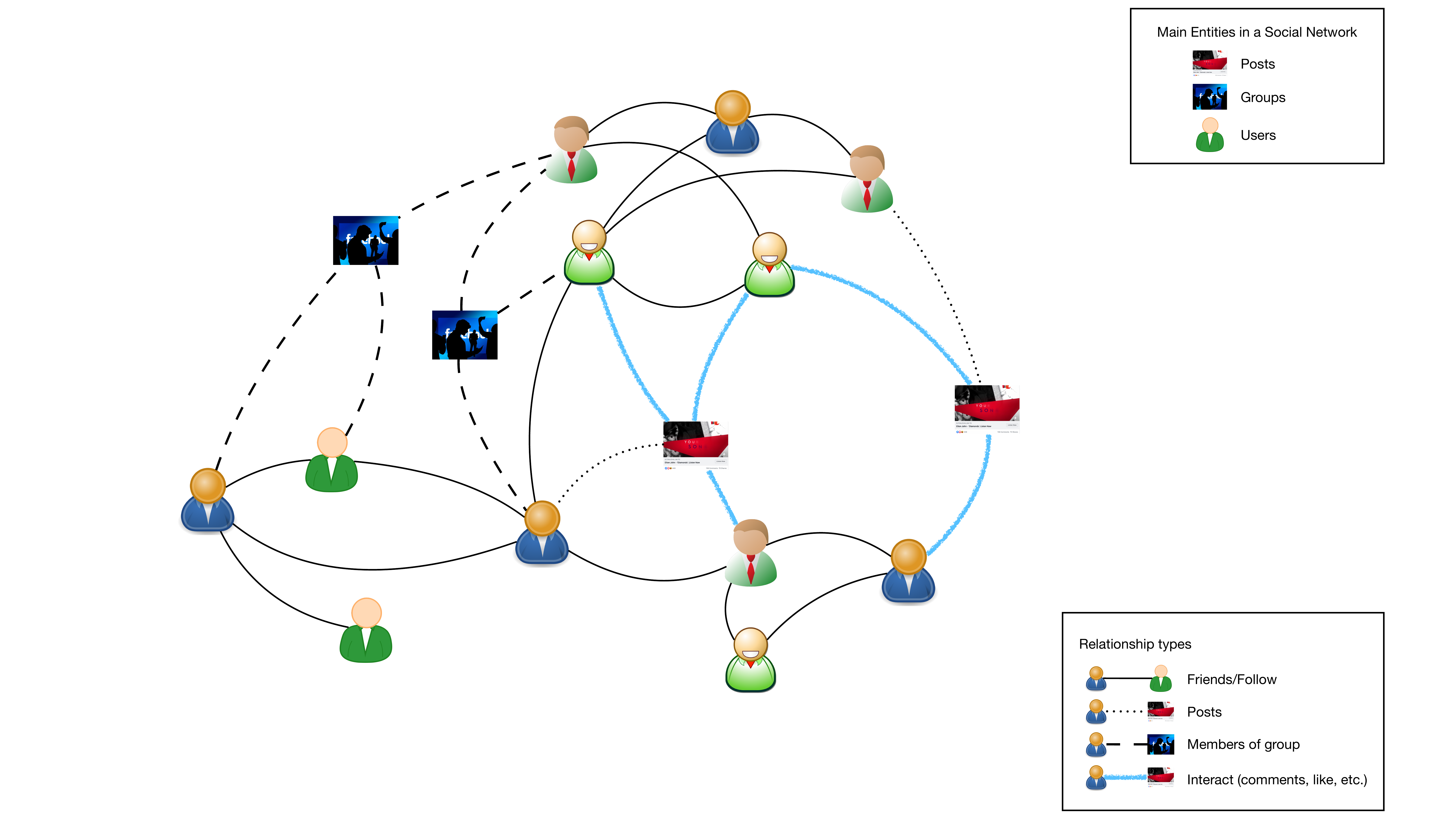}
\caption{The main components of a social networks: entities (users, posts, etc.), and their relationships.}
\label{fig:socialnetwork}
\end{center}
\end{figure}

\begin{figure}[ht]
\begin{center}
\includegraphics[width=3in]{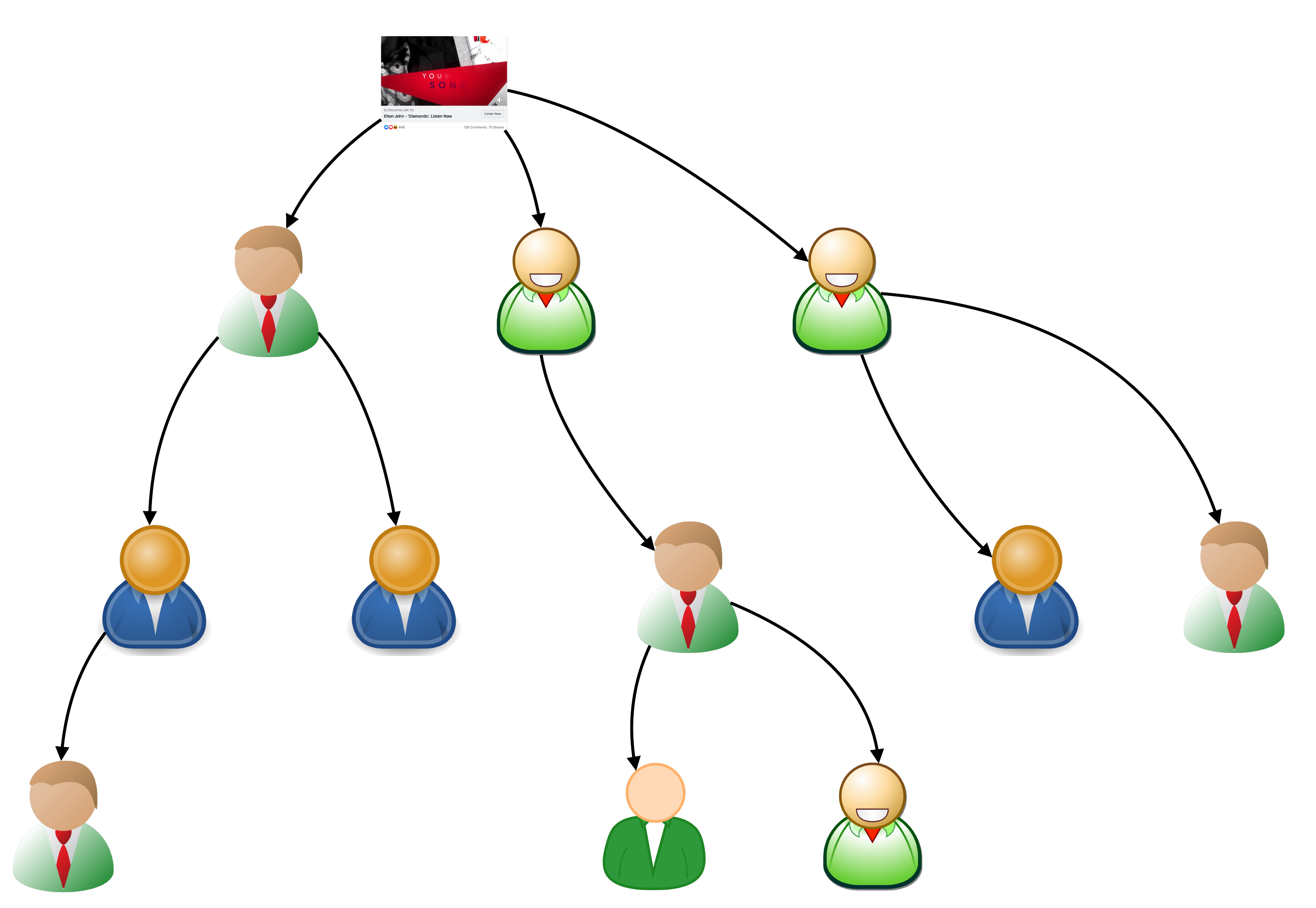}
\caption{The re-share activity on a social network forms a tree structure that is called an \textit{information cascade}. At the root of the cascade there is the original post and each arrow points to a user in the social network that shared it with their friends/followers. The patterns arising in an information cascade are used to infer integrity-related issues in the social network.}
\label{fig:cascade}
\end{center}
\end{figure}

As mentioned above, a common set of techniques consider graph representations of the network in order to leverage more distant effects. 
Wu and Huan~\cite{wu2018tracing}, considered analyzing network reactions in order to predict the spread of misinformation. The TraceMiner System that is designed to classify a piece of textual information as misinformation or not is based on the way it spreads on the social network. TraceMiner operates in two stages. In the first stage the system learns user embeddings based on the social graph. In addition, the system computes the {\em information cascade} generated from the re-share activity of posts (see Figure~\ref{fig:cascade}). In the second stage, the embeddings and the information cascades are fed into a sequence-based classifier that is trained on the task of detecting fake news.
  
More recent works apply Geometric Deep Learning (GDL) (i.e., deep learning on non-euclidean spaces such as graphs or manifolds) to the problem of detecting misinformation spread on the network~\cite{monti2019fake}. According to Monti {\em et al.}~\cite{monti2019fake}, the main benefit of GDL techniques is that they are agnostic to the kind of features you design on graphs or even the kind of data you consider. In a sense, GDL can be seen as a technique that allows to holistically incorporate heterogeneous data such as information about users, content, and the network structure. In terms of the graph shown in Figure~\ref{fig:socialnetwork}, this means allowing more types of nodes and edges between nodes. 
   
More formally, Monti {\em et al.}\ define the problem as that of classifying a URL as hoax or not. They built an information cascade graph (Figure~\ref{fig:cascade}), $G_u$, out of the interactions that the url, $u$, has had in the social network. The graph contains nodes for users who have shared $u$ and the re-sharing patterns. Two users, $i$ and $j$, are connected by a direct edge $i \rightarrow j$ if $i$ re-shared the URL $u$ from $j$. Nodes and edges in $G_u$ are also labelled with a set of features representing the users (node labels) and the type of interactions (edge labels). The graph, $G_u$, is then fed into a graph convolutional neural network for the classification task. 
(Note that the same graph can be used to classify a single URL or the set of re-shares for that URL). Monti {\em et al.}\ conduct experiments on a large set of annotated fake and true stories spread on Twitter. The model achieves very high accuracy, nearly 93\% ROC AUC. One of the important benefits of this work is that the model allows for detecting fake stories within a few hours after they have been posted.

Tambuscio {\em et al.}~\cite{tambuscio2015fact} leverage techniques from the analysis of disease spread to propose a theoretical model for the problem of detecting misinformation. Their  analysis is also based on the concept of information cascades.  The two main parameters they consider are the probability that a news mentioned in a tweet is correct (i.e., fact checked), and the probability of someone retweeting a news. The latter corresponds to the probability of the misinformation bit to be spread. The interesting finding shown in the paper is that when the probability of correctness is above a threshold defined by network parameters, then the probability of a hoax to spread through the network goes to zero. 
   

\subsection{Understanding actors on the network}
\label{section:actors}

Understanding actors on a social network and the connections between them can be used to build signals that can increase the effectiveness of integrity violation classifiers and even catch an entire set of violations at once. Similarly, understanding the relationship between two individuals on a social network might help to clarify ambiguous cases. For example, words or terms that might otherwise be violating can be used self-referentially or in an empowering way between two friends exchanging comments on a private post.
To understand actors, researchers have considered several types of information, including the actors' past behavior and interests, their connection to other actors and the communities to which they belong. This body of work traces back to early work since the inception of social networks that considered social context for content classification~\cite{mccallum2005topic, otte2002social, fortuna2007improving, stone2008autotagging}.

Most of the work proposed in the literature focuses on analyzing the network formed by reactions (share, comments, reactions) to posts that users make.
Some of the most effective works have considered automatically extracting user embeddings,  by running graph learning algorithms such as DeepWalk~\cite{perozzi2014deepwalk}, Line~\cite{tang2015line} and Node2Vec~\cite{grover16node2vec}, on those networks.  Those embeddings are then used in downstream tasks such as that of detecting misinformation~\cite{wu2018tracing} or hate speech. In a system that was used in practice in several integrity tasks at Facebook, Noorshams {\em et al.}~\cite{noorshams2020ties} built a model that enriches the user embeddings with a temporal model of reactions to a post.

Mishra \textit{et al.}~\cite{mishra2018author}, and subsequently Del Tredici \emph{et al.}~\cite{del2019you}, considered the problem of understanding actors in the context of abusive language classifiers. Mishra \textit{et al.}~\cite{mishra2018author} incorporate a user's embedding as features in a downstream abusive language classifiers, while Del Tredici \emph{et al.}~\cite{del2019you} used Graph Attention Network (GAT), a Geometric Learning method, to train a classifier for hate speech directly on the social network structure shown in Figure~\ref{fig:socialnetwork}.

In addition to detecting hate speech and misinformation, the analysis of the relationships in the network have also been used in other integrity problems, such as preventing suicides~\cite{mishra2019snap} and detecting controversial discussions~\cite{jang2018explaining}.  There has also been quite a bit of work that considered {\em only} the network structure, usually using complex network analysis~\cite{menczer2020first}, for detecting integrity violations. 

\medskip
\noindent{\bf Coordinated action:}
Coordinated actions among multiple actors have become a common strategy to promote integrity violating content with possibly severe consequences. Coordinated behaviors refer to sets of actors that together try to ensure that a post gets wider distribution and appears more authoritative. For example, actors may band together to spread misinformation that election polling locations have closed earlier to prevent people from voting.  Often, these coordinated actions are achieved through social bots. 

Shao {\em et al.}~\cite{shao18-nature} showed evidence that social bots played a disproportionate role in spreading articles from sources with low credibility.  Bovet and Makse~\cite{bovet19-nature} showed that social bots likely played a major role in spreading political misinformation on Twitter during the 2016 election in the United States.  One of the techniques used by the bots to raise the visibility of questionable news was to reply to the posts and and mention influential users in the replies. In contrast, Vosoughi {\em et al.}~\cite{roy18-science} came to a different conclusion. They tested the speed of propagation of fake news in the presence or absence of automated social bots and concluded that the virality of a piece of misinformation is not affected by the presence or absence of social bots. 

A better understanding of the  effect of coordinated action is clearly an area for future research. In one recent work on the topic, Pacheco {\em et al.}~\cite{pacheco2020uncovering} propose a method for exploiting the network structure to uncover coordinated activities within a social network. Their method, Coordination Detection Framework, works by detecting the surprising lack of independence of actions by users and exploiting this analysis to cluster together users that are likely to coordinate against some targets. Tracing the behavior of the users, the method discovers groups of users that are connected to the same shared resource (e.g., URLs), groups of users in the same geographical location, or a combination thereof. The method then uses this information to cluster users in groups that are likely to be doing malicious activities. 

\medskip
\noindent
{\bf Outlook:} In addition to the open issues mentioned above, another research challenge is to develop a holistic approach that understands a multitude of integrity violations. The graph-based methods proposed so far have focused on a single violation at a time. However, the graph model itself can support a combination of different violation-detection tasks. The potential benefit lies in the hypothesis that models learned from one violation can transfer to other violations. 

\section{Emerging topics and challenges}
\label{section:others}

The challenges in preserving integrity are constantly shifting due to the adversarial nature of the problem and because the policy and technology backdrops in which it operates are dynamic. In this section we touch upon challenges that arise from more recent developments in this environment and some approaches for addressing them.\footnote{In Figure~\ref{fig:marijuana}, the image on the left is fried broccoli and the image on the right is marijuana.}

\subsection{Integrity while maintaining privacy}
\label{section:privacy}
Maintaining user privacy during online interactions is an important and rising concern. In response to user sentiment about privacy, most messaging applications are moving to be encrypted from end to end. As a consequence, since the content of the messages is no longer visible to the service provider, any analysis of content to detect integrity violations needs to be performed on the device itself. In this section we consider some of the challenges associated with enforcing integrity in this new setting and approaches that have been proposed in the literature.

\paragraph{On device media matching:} One approach to detecting violating content that is known to authorities is based on non-reversible hashes. In this setup, the client sends a non-reversible hash of the content (e.g., an image) to the server to be checked against a bank of known violations. If the hash does not match the database, then the contents of the data are still private. Several techniques have been proposed for such hashes in the context of child sexual abuse material (CSAM) or terrorism, including PhotoDNA~\cite{photodna, Singh_2019_CVPR_Workshops} and PDQ-Hash~\cite{fb_pdq}.

However, even this seemingly secure method has some drawbacks. Specifically, the service provider can check the image hash coming from the device against {\em any} content, not only the known violations. For example, if the service provider is run by a government that is trying to identify dissenters, they can try to identify individuals sending certain political memes. This scenario brings up the question of auditability and transparency of hash lists to ensure that the technology is only used for the intended purpose. However, in the context of CSAM this can be tricky because sharing violating content is illegal.

An alternative approach is based on sending image hashes for the known violations to the devices and securing the image hashes so that the list doesn't leak to bad actors. Enforcement can then be done on the device when the content arrives, with no data leaving the device. Of course, as with all privacy approaches, it is also important to consider whether the techniques lead to a {\em perceived} sense of privacy by naive users. 

\paragraph{On device enforcement:} An alternative approach that applies to other integrity violations is to perform the inference completely on the device.  The machine learning model that is used for the inference needs to be trained offline on publicly available datasets and then shipped to devices. On-device inference is limited by the memory and processing power of the device as well as potentially having an adverse effect on its battery life. These limitations determine the number and capacity of the models one can deploy. In the context of text analysis, storing the word embedding table is the bottleneck in terms of memory space. Several works have proposed techniques for addressing this challenge using variations of hashing-based models that don't require lookup tables~\cite{DBLP:journals/corr/JoulinGBDJM16,DBLP:conf/emnlp/RaviK18}.

In addition to performing inference on the device, it is conceivable to train the models on the device. On device training has been used for learning models for keyboard and query suggestions, where it is highly important to have personalized models. However, in the context of integrity, we are most interested in the problem of learning a global model that captures violations. In the context of integrity, certain kinds of violations are very rare, so even if we use techniques like secure aggregation~\cite{DBLP:conf/ccs/BonawitzIKMMPRS17} to aggregate training data from multiple devices, we may still see entire cohorts of users without positive examples of certain kinds of violations. 

\paragraph{Measurement:} 
Even though on-device media match and on-device inference offer viable approaches to detecting violating content, measuring the success of the algorithms remains a significant concern.
In particular, the effectiveness of a model trained offline and shipped to clients will typically degrade over time, and if its performance cannot measured, it would be hard to know when it needs to be fixed.   

Differential privacy~\cite{DBLP:conf/tamc/Dwork08,DBLP:conf/ccs/AbadiCGMMT016} provides one approach to this problem.  In a nutshell, differential privacy allows a central server to collect data on sensitive topics while preserving confidentiality. The technique relies on adding noise to the data so that the user retains strong deniability. One popular technique that builds on top of differential privacy is Rappor~\cite{DBLP:conf/ccs/ErlingssonPK14}, which enables collecting statistics in a privacy-preserving manner and is suitable for deploying at scale, even in low-resource environments. 

However, in the context of preserving integrity, there is one more challenge that needs to be addressed. One of the biggest downsides of adding noise to data in differential privacy is that it destroys the information from rare events. In~\cite{DBLP:conf/csfw/HsuGHKNPR14} it is shown that for a balanced privacy level epsilon of ln(3), to detect events with frequency 0.1\% 100 million samples are required. For events with frequency 0.01\%, this goes up to 10 billion samples. Hence, for this privacy level it is unlikely that one will be able to use RAPPOR-like methods to measure the prevalence of many integrity violations such as CSAM. On the other hand, there is some potential to use Rappor for higher prevalence problems such as misinformation campaigns.

\subsection{Management of human content review workforce}
\label{section:humans}

One of the critical steps in enforcing integrity is inspection of content by human content reviewers. The main role of the review workforce is to recognize content that violates the integrity policies. The content that is fed into the worker pipeline originates either from an enforcement system or from users reporting violations. The labels that these workers produce are crucial to training the machine learning models that are used by the enforcement systems. In parallel, there is a set of content reviewers who  work for third-party organizations and fact-check content suspected to be misinformation.

One challenge with operating such a workforce is the high volume of content that needs to pass their judgment. In addition, we need to be able to adapt and respond to scenarios when there is a sudden drop in workforce availability, as has happened during the COVID-19 pandemic. The well-being of the content reviewers is a critical concern--the content can be graphic or otherwise objectionable.  Operators of content moderation work forces may put several resources in place, such as 
providing access to licensed counselors, providing group therapy sessions, and screening applicants for suitability for the role as part of the recruiting process.

One of the key techniques that is used to make the worker pipeline more efficient is to group near-duplicate content, using techniques mentioned in Section~\ref{section:content}, such as~\cite{photodna,DBLP:journals/corr/ToliasSJ15}. For labeling videos, an important time-saving technique is to try to localize where in the video there might be violating content so that the human doesn't need to view the entire video. Techniques such as~\cite{DBLP:conf/iccv/SelvarajuCDVPB17} are commonly used in practice. 

A more subtle challenge is the potential for bias~\cite{DBLP:journals/debu/Halevy19} when making decisions about claims that may be subjective.  As an example, as a member of a particular ethnic group, I may be more likely to label a joke about that group as hate speech. In the case of misinformation companies engage with neutral fact checkers to address the potential for bias. Facebook, for example, partners with third-party fact checkers that are certified through a non-partisan International Fact Checking Network (IFCN) and follow IFCN's Code of Principles that include nonpartisanship and fairness; transparency of sources; transparency of funding and organization; transparency of methodology; open and honest corrections policy.

\subsection{Preparing for changes}
The integrity landscape changes rapidly.  While entirely new types of violations do arise, the more common challenge is that new topics become popular and therefore the subject of violating content, especially for misinformation and hate speech. For example, during the debate about migrants at the southern border of the U.S.\ in 2019, new hate speech was targeted at migrants using the topic of ``the caravan'', or more recently COVID-19 generated hate speech towards Asians. In doing so, violators use new words that are unknown to the enforcement systems, e.g., ``caravandals''. Programmatic labeling of data can be a useful tool in this context~\cite{DBLP:conf/sigmod/BachRLLSXSRHAKR19}. With programmatic labeling, we can quickly use knowledge we have about the domain to generate some noisy training data for our classifiers, or at least generate a dataset with a better balance between positive and negative examples that can then be given to a human workforce. 
 
Policy changes also require to potentially re-train 
multiple classifiers. For example, the definition of what constitutes political ads is often in flux. If the policy changes, an advertisement for solar panels or an advertisement from a non-political group calling for action on climate change may be classified differently than they were before.  Given the complex systems like the ones driving integrity decisions in social media, fast re-training of classifiers is a challenge. In such situations, we may have to face the choice of being slow to implement policy changes, be fast but make mistakes, or over-use a human review task force to do a better classification job.

\section{Conclusions}
\label{section:conclusions}

Given the prominence of social media as a medium for sharing news and information, protecting integrity of online content is of prime societal importance. As this survey has shown, protecting integrity operates within a rapidly changing landscape because of policy and societal changes as well as adversaries becoming more sophisticated. From the technological perspective, preserving integrity presents a challenge that pushes on the boundaries of many aspects of Artificial Intelligence and its adjoining fields. We have also touched on the need for greater collaboration between industry and academia. The survey outlined some of the techniques that have shown promise in practice and require more attention from academia, and we have argued for creating more datasets that would foster research on integrity topics.

An important issue that integrity will soon grapple with is the boundary between content that obviously violates the policy guidelines and borderline content that may offend a wide audience. The overarching goal of social media is to increase the number of positive experiences that users have on the platform and minimize the negative experiences. Borderline content is a major source of such negative experiences. However, determining whether a piece of content will lead to a negative experience for a user is a highly subjective call and may be perceived as too much interference by the social media platform. Treading this fine line in a healthy fashion will surely be an important challenge in the upcoming years.

 \bibliographystyle{ACM-Reference-Format}
  \bibliography{references-modified}

\end{document}